\documentclass[aps,pra,12pt, showpacs,showkeys,superscriptaddress,longbibliography]{revtex4-1}% floatfix,
\pdfoutput=1
\usepackage{amsmath,amsfonts,amsthm}

\usepackage[utf8]{inputenc}
\usepackage[english]{babel}

\usepackage{graphicx}
\usepackage{verbatim}
\usepackage{tensor}
\usepackage[all,cmtip]{xy}
\usepackage{dsfont}

\usepackage{color}
\definecolor{dred}{rgb}{.8,0.2,.2}
\definecolor{ddred}{rgb}{.8,0.5,.5}
\definecolor{dblue}{rgb}{.2,0.2,.8}
\usepackage{hyperref}
\hypersetup{
  pdftitle={A diagrammatic approach to map-state and channel-state dualities},
  pdfauthor={Bergholm, Ville},
  pdfsubject={quantum circuits, tensor network states},
  pdfkeywords={quantum circuit model, quantum information, tensor network,
  categorical quantum mechanics, channel-state duality, map-state duality}
}

\newcommand{\inprod}[2]{\ensuremath{\left\langle #1, #2 \right\rangle}}
\newcommand{\inprodHS}[2]{\ensuremath{\left\langle #1, #2 \right\rangle}} % _\text{HS}
\newcommand{\norm}[1]{\ensuremath{\left\| #1 \right\|}}
\newcommand{\bra}[1]{\ensuremath{\langle #1|}}
\newcommand{\ket}[1]{\ensuremath{|#1\rangle}}
\newcommand{\braket}[2]{\mbox{$\langle #1|#2\rangle$}}
\newcommand{\ketbra}[2]{\mbox{$|#1\rangle\langle #2|$}}

\newcommand{\eq}{\iff} %{\Leftrightarrow}
\DeclareMathOperator{\Rank}{rank}
\DeclareMathOperator{\Tr}{Tr}

\DeclareMathOperator{\Span}{span}

\DeclareMathOperator{\Ker}{Ker} % kernel
\DeclareMathOperator{\Hom}{Hom} % homomorphisms
\DeclareMathOperator{\End}{End} % endomorphisms
\newcommand{\ie}{i.e.}

\newcommand{\gate}[1]{\ensuremath{\text{\sc #1}}}

\newcommand{\SWAP}{\gate{SWAP}}

\newcommand{\opstateup}[1]{#1}
\newcommand{\opstatedown}[1]{{#1}^T}

\newcommand{\supop}[1]{\widetilde{#1}}
\newcommand{\choi}[1]{\widehat{#1}}

\newlength{\figtextvoffset}
\newcommand{\fscale}{1.0}

\newcommand{\fhelp}[1]{\scalebox{\fscale}{$#1$}}
\newcommand{\f}[1]{\settoheight{\figtextvoffset}{\fhelp{#1}}\raisebox{-0.5\figtextvoffset}{\fhelp{#1}}}
\newcommand{\s}[1]{\scriptsize{$#1$}}

\newcommand{\COPY}[2]{\gate{COPY}\ensuremath{_{#1}^{#2}}}

\newcommand{\I}{\mathds{1}}   % Hilbert space identity op (\openone is ugly)
\newcommand{\eye}{\mathbf{1}} % unit for tensor product

\newcommand{\Figref}[1]{Fig.~\ref{#1}}

\newcommand{\be}{\begin{equation}}
\newcommand{\ee}{\end{equation}}

\newcommand{\hilbert}[1]{\ensuremath{\mathcal{#1}}}

\newcommand{\cupket}[1]{\ket{\supset}_{#1 \otimes #1}}
\newcommand{\capbra}[1]{\bra{\subset}_{#1 \otimes #1}}
\newcommand{\cuprho}{\ket{\supset}\bra{\subset}}
\newcommand{\cupk}{\ket{\supset}}

\newcommand{\obs}{x} % observable
\newcommand{\dualc}[1]{{#1}^*} % dual of a channel
\newcommand{\cc}[1]{\overline{#1}} % complex conjugate

\newcommand{\isom}{\cong}  % isomorphic to

\theoremstyle{plain}
\newtheorem{theorem}{Theorem}
\newtheorem{proposition}[theorem]{Proposition}
\newtheorem{lemma}[theorem]{Lemma}

\theoremstyle{definition}
\newtheorem{definition}{Definition}
\newtheorem{example}[theorem]{Example}

\newcommand{\HH}{\hilbert{H}}
\newcommand{\R}{{\mathbb R}} % reals
\newcommand{\C}{{\mathbb C}} % complex numbers
\begin{document}
\title{A diagrammatic approach to map-state and channel-state dualities}
\author{Ville Bergholm}
\email{ville.bergholm@iki.fi}
\affiliation{ISI Foundation, Via Alassio 11/c, 10126 Torino, Italy}

% alg methods in qm 03.65.Fd
% qm formalism 03.65.Ca
% q systems with finite H space 03.65.Aa

\begin{abstract}
Diagrammatic representation and manipulation of tensor networks has
proven to be a useful tool in mathematics, physics, and computer science.
Here we present several important and mostly well-known theorems 
regarding the dualities between linear maps and bipartite pure quantum states,
and the dualities between quantum channels and bipartite mixed quantum states, in diagrammatic form.
The graphical presentation makes the proofs very compact and in some cases even
intuitive.
\end{abstract}
\maketitle

%\tableofcontents

\section{Introduction}
\label{sec:intro}
%A tensor network is a set of tensors where some of the indices have
%been contracted. Any uncontracted index is called an ``open leg'' of
%the network, whereas 
%They can be used to represent...
%computational tools for describing quantum many-body systems

Diagrammatic methods for describing tensors and their connections have
a long history in physics, mathematics and computer science.
Their importance stems from the fact that they enable one to
perform mathematical reasoning and even actual calculations using
intuitive graphical objects instead of abstract mathematical entities.
In the early 1970s Penrose introduced a somewhat informal but expressive
graphical notation for representing tensor network expressions such as
the ones one encounters in general relativity~\cite{Penrose_string_diagram}.
Perhaps the first one to note the importance of diagrammatic methods in
quantum information science was David Deutsch.
Today quantum circuit diagrams (QCDs), a well-defined subclass of
tensor network diagrams, are a standard tool for
describing quantum algorithms and protocols.

The channel-state duality, or Choi-Jamiołkowski isomorphism, is a
central result in quantum information science. It establishes a direct one-to-one
correspondence between quantum channels (processes that map valid quantum
states to valid quantum states), and mixed quantum
states on a larger Hilbert space.
In this work we aim to provide a complementary viewpoint to the usual abstract
algebraic derivation and presentation of these results.
The graphical presentation tends to make the proofs very compact, easy to
follow, and in some cases even intuitive.
Our presentation occasionally follows~\cite{arrighi2004} and~\cite{vv2003}.
Related results can be also found in~\cite{wood2011}.
%\cite{johnson2014}.

We will start by introducing the basic string diagram notation for tensor networks
in Sec.~\ref{sec:notation}. Our notation bears some resemblance to
quantum circuit diagrams, but additionally borrows ideas from the
graphical language used to describe symmetric monoidal categories~\cite{Selinger09}.
Our presentation of the string diagram techniques is necessarily
somewhat cursory. For a more complete treatment, see~\cite{BB10}.
Then, in Sec.~\ref{sec:opstate} we will introduce the basic map-state
duality and discuss some of its implications.
In Sec.~\ref{sec:channelop} we present the Choi-Jamiołkowski
isomorphism and derive several important and famous results concerning the
properties of quantum channels.
%We will conclude by the discussion in Sec.~\ref{sec:discussion}.
In the Appendices, we provide some relevant background on linear algebra~\ref{sec:linalg},
and quantum states~\ref{sec:states}.

Finally, a few words on nomenclature.
Using the terribly abbreviated physics vocabulary, when we say a
linear map or matrix is positive, we actually mean positive semidefinite.
% Ket means a vector (and not a Hilbert space ray).
In this work we deal with finite-dimensional vector spaces only.

\section{Notation}
\label{sec:notation}

Here we briefly explain the basic string diagram notation for tensor networks.
For a more complete treatment together with proofs, see~\cite{BB10}.

\begin{figure}
\renewcommand{\fscale}{1.0} \def\svgwidth{0.7\textwidth} 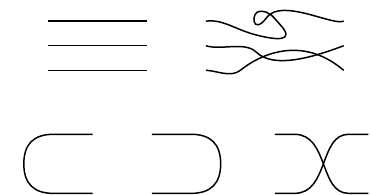
\caption{Basic wire structures.
  (a) Identity map $\I_{A \otimes B \otimes C}$, or the Hilbert space $A \otimes B \otimes C$.
  (b) Cup $\cupket{A}$.  (c) Cap $\capbra{A}$. (d) $\SWAP{}_{A \otimes B}$ gate.
\label{fig:basic-wiring}}
\end{figure}

Horizontal wires represent Hilbert spaces.
We assume that every Hilbert space $A$ comes with a preferred
\emph{computational basis} $\{\ket{k}_A\}_k$.
Stacking the wires vertically corresponds to a
system comprised of several subsystems, as shown in \Figref{fig:basic-wiring}a.
Unless the types are clear from the context, each wire should be explicitly labeled.
Unlike in standard QCDs, the wires are allowed to deviate from a
straight horizontal line as long as they do not cross each other, or
reverse direction. Soon, however, we will relax both of these rules
by introducing some additional wire-like diagram elements.

\begin{figure}
\renewcommand{\fscale}{1.0} \def\svgwidth{0.9\textwidth} 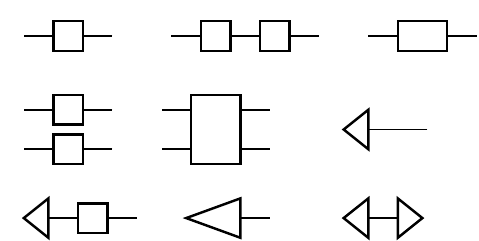
\caption{Linear maps and vectors. (a) Map $f: A \to B$. (b) Composition.
(c) Tensor product. (d) Vector $\ket{\psi} \in A$.
(e) Map acting on a vector. (f) Inner product $\inprod{\phi}{\psi}$.
\label{fig:gates}}
\end{figure}

Linear maps between Hilbert spaces are represented by labeled boxes,
with the domain wire(s) connecting to the left side of the box, and
the codomain wire(s) to the right side. Maps are concatenated simply
by placing the boxes next to each other on the wire.
Similarly, two boxes stacked atop each other represent the tensor product
of the corresponding maps, as shown in \Figref{fig:gates}a--c.
Consequently, a bare wire can also be understood as the identity map.

Vectors (or kets) are represented by labeled left-pointing triangles,
and the corresponding covectors (one-forms/functionals/bras, via Riesz's theorem)
as labeled right-pointing triangles. Maps operating on vectors, inner
products, and tensor products are represented in the obvious way,
as shown in \Figref{fig:gates}d--f.

Taking the Hermitian conjugate of an expression is accomplished by
mirroring the corresponding diagram in the left-right direction, and
adding a dagger symbol to each box label.
% This could be avoided e.g. by adding arrows to the boxes...
Now we will complete our notation by introducing three additional wire-like structures.

\begin{definition}[Cup and cap]
Given the computational basis, a
\emph{cup} is the diagram element that corresponds to the 
bipartite vector
\be
\cupket{A} := \sum_{i} \ket{i}_{A} \otimes \ket{i}_{A}.
%= \delta\indices{^{ij}} \ket{ij}_{A \otimes A}.
%\quad \text{where} \;
\ee
Likewise, a
\emph{cap} is the diagram element that corresponds to
the corresponding covector:
\be
\capbra{A} := \sum_{i} \bra{i}_{A} \otimes \bra{i}_{A}.
%= \delta\indices{_{ij}} \bra{ij}_{A \otimes A}.
\ee
The corresponding diagrams are presented in \Figref{fig:basic-wiring}bc.
Note that
normally the symbol inside the ket or bra does not change when
taking the dagger:
$\ket{\psi}^\dagger = \bra{\psi}$.
Here it does,
$\cupket{A}^\dagger = \capbra{A}$,
but only for the sake of aesthetics.
We essentially define two redundant symbols for the same thing:
$\cupket{A} = \ket{\subset}_{A\otimes A}$.
%\footnote{
%In a category-theoretical treatment, the cup and the cap correspond to the dagger-compact structures~$\eta_A$ and
%$\epsilon_A$, respectively, in the category of extended quantum circuits~\cite{BB10}.}
\end{definition}

It is easy to notice that interpreted as a pure quantum state, $\cupket{A}$ is proportional to
the maximally entangled generalized Bell state $B_{0,0}$, %in Def~\ref{def:bell},
\be
\ket{B_{0,0}}_{A \otimes A} =
\frac{1}{\sqrt{\dim A}} \cupket{A},
\ee
and that the other generalized Bell states are locally
equivalent to it.

\begin{definition}[\SWAP{} gate]
$\SWAP{}_{A \otimes B}: A \otimes B \to B \otimes A$
exchanges the order of two subsystems and is
consequently represented by crossing wires, see
\Figref{fig:basic-wiring}c.
\be
\SWAP{}_{A \otimes B} = \sum_{ab} \ket{ba}_{B \otimes A} \bra{ab}_{A \otimes B}.
\ee
In the case where $A=B$, we can instead interpret it as swapping the
\emph{states} of the two subsystems, thus recovering the
usual definition.
\end{definition}

\begin{figure}
\renewcommand{\fscale}{1.0} \def\svgwidth{0.95\textwidth} 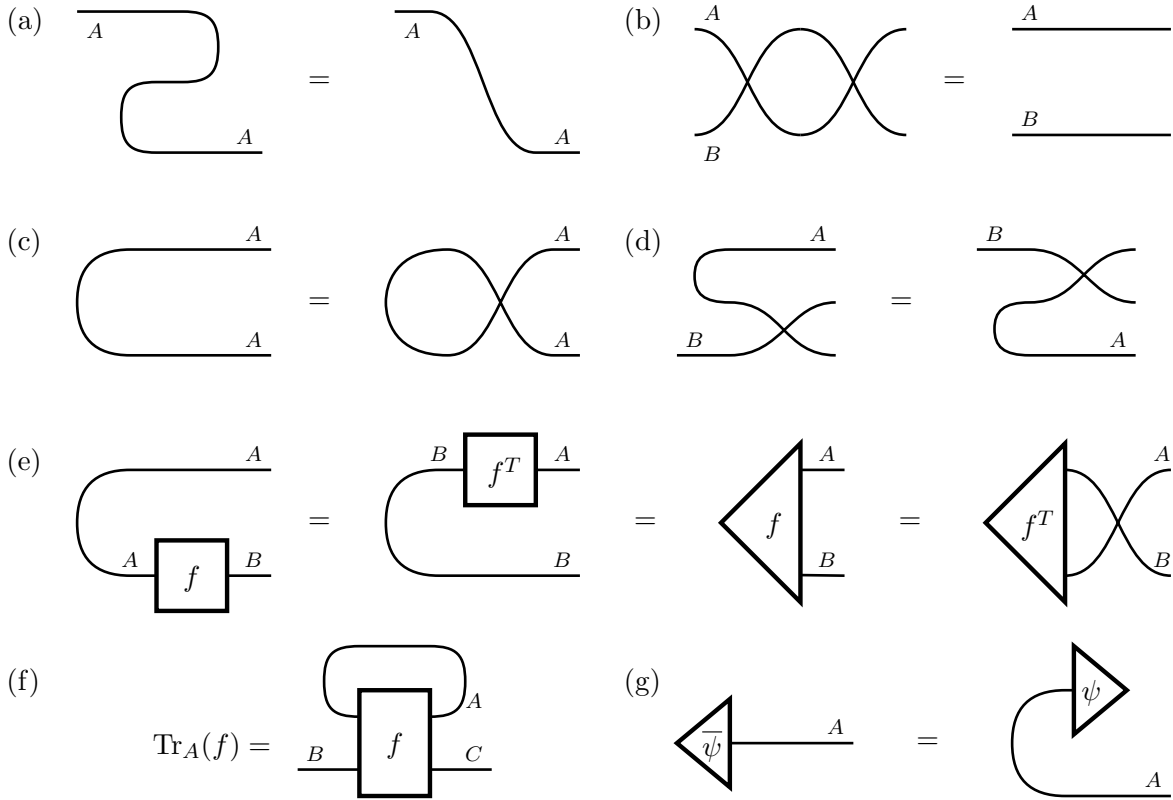
\caption{Wire identities.
(a) Snake equation.
(c) $\SWAP{}_{A \otimes B}^{-1} = \SWAP{}_{B \otimes A}$.
(c) Cup symmetry.
(d) Cup crossing a wire.
(e) ``Sliding'' an operator~$f$ around a cup (or cap) transposes it in the
computational basis. Definition of the operator state~$\ket{\opstateup{f}}$.
(f) Trace.
(g) Conjugate state.
\label{fig:cups}
}
\end{figure}

Now we will list (without proof) some fundamental properties of cups, caps and \SWAP{}s,
corresponding to the diagram identities in~\Figref{fig:cups}.

\begin{proposition}[Snake equation (\Figref{fig:cups}a)]
A cup and a cap can combine to cancel each other. In other words,
a double bend in a wire can be pulled straight.
\end{proposition}

\begin{proposition}[Inverse of \SWAP{} (\Figref{fig:cups}b)]
The inverse of $\SWAP{}_{A \otimes B}$ is simply $\SWAP{}_{B \otimes A}$.
Thus two stacked wires can be pushed through each other.
\end{proposition}

\begin{proposition}[Cup and cap symmetry (\Figref{fig:cups}c)]
Since the cup corresponds to a symmetric state, it immediately follows
that the relative order of the two subsystems is irrelevant. 
Diagrammatically this means the order of the outgoing wires can be swapped,
or a bend ``twisted'' into a loop.
\end{proposition} 

\begin{proposition}[Cup crossing a wire (\Figref{fig:cups}d)]
\SWAP{} interacts with a cup in the obvious way.
Hence a bend can be moved across a wire.
\end{proposition}

With these propositions, the interpretation of
\Figref{fig:basic-wiring}a becomes clear.
One can manipulate the wires almost as if they were rubber bands
confined in a two-dimensional plane without changing the meaning of
the diagram.
Mathematically the purpose of the various wire structures is to reorder and entangle subsystems in various ways.
When interpreting the diagrams physically, one can think of time flowing from left
to right, much like in ordinary QCDs, in which case the different
reshapes of a diagram correspond to different ways of obtaining the
same physical effect.

\begin{proposition}[Sliding operators around cups and caps; operator states
(\Figref{fig:cups}e)]~\\
\label{th:slide}
An operator~$f: A \to B$ can be moved (``slid'')
around a cup or a cap by transposing it in the computational basis.
Alternatively, there is a trivial isomorphism between a cup
followed by the operator~$f$ on the second subsystem, the
vector~$\ket{\opstateup{f}}$, % $:= \text{vec}(f)^k \ket{k}_{A \otimes B}$,
and a cup followed by the operator~$f^T$ on the first subsystem.
\end{proposition}
Note that $\ket{\opstateup{f}}$ is obtained by taking the matrix representation
of $f$ in the computational basis and
rearranging it column by column, left to right, into a column vector,
much like the $\text{vec}$ operation used in numerical software.

\begin{proposition}[Trace (\Figref{fig:cups}f)]
\label{th:trace}
Looping a wire from an output to an input of an operator~$f$ corresponds to taking a partial
trace of~$f$. Consequently, looping all the outputs into corresponding inputs
corresponds to the full trace.
\end{proposition}

\begin{proposition}[Conjugate states (\Figref{fig:cups}g)]
Cups and caps, together with the dagger functor, induce an isomorphism between states
$\ket{\psi}$ and their
\emph{conjugate states} $\ket{\cc{\psi}}$,
obtained by complex conjugating
the coefficients of the state in the computational basis.
\end{proposition}

At first it might seem strange that we should encounter basis-dependent
operations such as transposition and complex conjugation. However,
this is a direct result of us having chosen a preferred computational basis
and defined the cup and cap structures in terms of it.

\section{Map-state duality}
\label{sec:opstate}

We will now explore the cup-induced isomorphism (introduced in Proposition~\ref{th:slide})
between linear operators~$f: A \to B$
and bipartite vectors $\ket{\opstateup{f}}_{A \otimes B}$,
which may be interpreted as pure quantum states.
This is equivalent to the common \emph{vec}-mapping, or columnwise
vectorization of a matrix.
One immediate result is that all matrix/operator decompositions yield a
corresponding bipartite state decomposition, and vice versa.

\begin{figure}[h!]
\renewcommand{\fscale}{1.0} \def\svgwidth{0.9\textwidth} 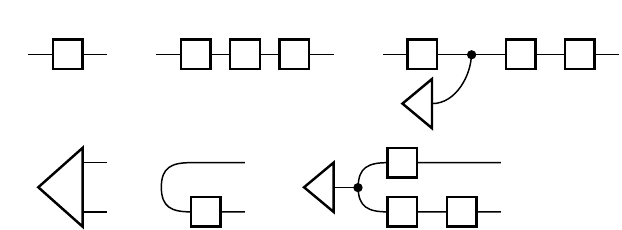
\caption{Correspondence between SVD and Schmidt decomposition.
(a) SVD of $f: A \to B$.
(b) Schmidt decomposition of~$\ket{\opstateup{f}}$.
}
\label{fig:SVD-Schmidt}
\end{figure}

\begin{proposition}[SVD $\eq$ Schmidt decomposition]
\label{th:SVD-Schmidt}
The singular value decomposition (SVD) of an operator $f: A \to B$
is equivalent to the Schmidt decomposition of~$\ket{\opstateup{f}}$, 
with the singular values $\{\sigma_i\}_{i=0}^{d-1}$
of $f$ corresponding to
the Schmidt coefficients of~$\ket{\opstateup{f}}$, as shown in~\Figref{fig:SVD-Schmidt}.
$d := \min(\dim A, \dim B)$.
\begin{proof}
The SVD of $f$ is
\be
\notag
f = U \Sigma V,
\ee
where $U: B \to B$ and $V: A \to A$ are unitary operators and
$\Sigma: A \to B$ is diagonal in the computational basis, with the
(real, nonnegative) singular values of $f$
on the diagonal.
% NOTE \{\sigma_i^2\} = \sigma(f^\dagger f)
$\Sigma$ can be represented diagrammatically as an order
$(1,1)$ diagonal dot (ddot), preceded or followed by a dimension-changing
ones-on-diagonal operator~$Q$ if $\dim A \neq \dim B$.
(Without loss of generality we may assume that $\dim A \le \dim B$.)

The ddot can further be split into a \COPY{}{2 \to 1} dot and a
unique ket $\ket{\sigma}$ 
% \order $(0,1)$ ddot
with nonnegative coefficients:
%The tensor $Q_{12}$ is only necessary if $\hilb{H}_1$ and $\hilb{H}_2$
%have different dimension.
\be 
\notag
\Sigma
= \sum_{j=0}^{d-1} \sigma_j \ket{j}_B \bra{j}_A
= \underbrace{\sum_{i=0}^{d-1} \ket{i}_B \bra{i}_A}_{Q_{AB}}
\underbrace{\sum_{j} \ket{j}_A\bra{jj}_A}_{\COPY{}{2 \to 1}}
\underbrace{\sum_k \sigma_k \ket{k}_A}_{\ket{\sigma}}
\qquad (\text{where} \: \sigma_k \ge 0).
\ee
The diagrammatic SVD is presented in~\Figref{fig:SVD-Schmidt}a.
By applying a cup to the SVD diagram 
we obtain~$\ket{\opstateup{f}}$.
We then slide $V$ around the bend and bend the ddot leg forward, arriving
at a diagrammatic representation for the Schmidt decomposition of
$\ket{\opstateup{f}}$,
\be
\notag
\ket{\opstateup{f}} = \sum_{k=0}^{d-1} \sigma_k V\indices{_k ^i} U\indices{^j _k} \ket{ij},
\ee
shown in~\Figref{fig:SVD-Schmidt}b.

Conversely, given a bipartite state $\ket{\psi}$ we may apply the
snake equation, apply the diagrammatic SVD and then perform the steps
above to obtain the diagrammatic Schmidt decomposition of~$\ket{\psi}$.
See also~\cite{BBL2012} for some applications.
\end{proof}
\end{proposition}

\begin{proposition}[Spectral decomposition $\eq$ conjugate state decomposition]
\label{th:spectral_dec}
If the operator $f: A \to A$ is normal, \ie{} $f^\dagger f = f
f^\dagger$, we may write its spectral (eigenvalue) decomposition as
\be
\notag
f = U \Lambda U^\dagger = \sum_{k=0}^{d-1} \lambda_k \ketbra{\lambda_k}{\lambda_k},
\ee
where $U: A \to A$ is unitary and $\Lambda: A \to A$ is diagonal,
and $U = \ketbra{\lambda_k}{k}$.
Much like in Proposition~\ref{th:SVD-Schmidt}, we may
write this in diagram form, with $\ket{\sigma}$ replaced
by $\ket{\lambda}$, holding the spectrum of~$f$.
Likewise, we may then find the corresponding
decomposition for~$\ket{\opstateup{f}}$:
\be
\notag
\ket{\opstateup{f}} = \sum_{k=0}^{d-1} \lambda_k \cc{U\indices{^i _k}} U\indices{^j _k} \ket{ij}
= \sum_{k=0}^{d-1} \lambda_k \ket{\cc{\lambda_k}} \ket{\lambda_k}.
\ee

In the special case where $f$ is hermitian, the spectrum is
real. Furthermore, if $f \ge 0$, the spectrum is nonnegative as well.
\end{proposition}

\begin{proposition}[Purification of positive operators]
\label{th:purification}
Any positive operator~$\rho: B \to B$ can be \emph{purified}, \ie{}
represented as the partial trace of a bipartite positive rank-1
operator
\mbox{$\ketbra{\opstateup{f}}{\opstateup{f}} \in \End(A \otimes B)$}:
\be
\notag
\rho = f f^\dagger = \Tr_A(\ketbra{\opstateup{f}}{\opstateup{f}}),
\ee
where $f: A \to B$, for any
$\dim A \ge \Rank \rho$. % \le \dim B$.
Conversely, for every nonvanishing bipartite vector~$\ket{f} \in A \otimes B$ the expression
$\Tr_A(\ketbra{f}{f})$ gives a positive operator of
rank~$\le \min(\dim A, \dim B)$.
\begin{proof}
See the diagram identity below:
\be
\notag
\renewcommand{\fscale}{1.0} \def\svgwidth{0.9\textwidth} %% Creator: Inkscape inkscape 0.48.4, www.inkscape.org
%% PDF/EPS/PS + LaTeX output extension by Johan Engelen, 2010
%% Accompanies image file 'purification.pdf' (pdf, eps, ps)
%%
%% To include the image in your LaTeX document, write
%%   \input{<filename>.pdf_tex}
%%  instead of
%%   \includegraphics{<filename>.pdf}
%% To scale the image, write
%%   \def\svgwidth{<desired width>}
%%   \input{<filename>.pdf_tex}
%%  instead of
%%   \includegraphics[width=<desired width>]{<filename>.pdf}
%%
%% Images with a different path to the parent latex file can
%% be accessed with the `import' package (which may need to be
%% installed) using
%%   \usepackage{import}
%% in the preamble, and then including the image with
%%   \import{<path to file>}{<filename>.pdf_tex}
%% Alternatively, one can specify
%%   \graphicspath{{<path to file>/}}
%% 
%% For more information, please see info/svg-inkscape on CTAN:
%%   http://tug.ctan.org/tex-archive/info/svg-inkscape
%%
\begingroup%
  \makeatletter%
  \providecommand\color[2][]{%
    \errmessage{(Inkscape) Color is used for the text in Inkscape, but the package 'color.sty' is not loaded}%
    \renewcommand\color[2][]{}%
  }%
  \providecommand\transparent[1]{%
    \errmessage{(Inkscape) Transparency is used (non-zero) for the text in Inkscape, but the package 'transparent.sty' is not loaded}%
    \renewcommand\transparent[1]{}%
  }%
  \providecommand\rotatebox[2]{#2}%
  \ifx\svgwidth\undefined%
    \setlength{\unitlength}{233.43811417bp}%
    \ifx\svgscale\undefined%
      \relax%
    \else%
      \setlength{\unitlength}{\unitlength * \real{\svgscale}}%
    \fi%
  \else%
    \setlength{\unitlength}{\svgwidth}%
  \fi%
  \global\let\svgwidth\undefined%
  \global\let\svgscale\undefined%
  \makeatother%
  \begin{picture}(1,0.1923422)%
    \put(0,0){\includegraphics[width=\unitlength]{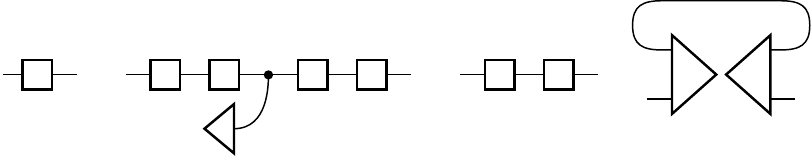}}%
    \put(0.81083017,0.13679055){\color[rgb]{0,0,0}\makebox(0,0)[b]{\smash{\s{A}}}}%
    \put(0.81083017,0.07607535){\color[rgb]{0,0,0}\makebox(0,0)[b]{\smash{\s{B}}}}%
    \put(0.96868953,0.13679055){\color[rgb]{0,0,0}\makebox(0,0)[b]{\smash{\s{A}}}}%
    \put(0.96868953,0.07607535){\color[rgb]{0,0,0}\makebox(0,0)[b]{\smash{\s{B}}}}%
    \put(0.84725928,0.1003616){\color[rgb]{0,0,0}\makebox(0,0)[b]{\smash{\f{f}}}}%
    \put(0.93226047,0.1003616){\color[rgb]{0,0,0}\makebox(0,0)[b]{\smash{\f{f}}}}%
    \put(0.58011259,0.10643316){\color[rgb]{0,0,0}\makebox(0,0)[b]{\smash{\s{B}}}}%
    \put(0.72582893,0.10643316){\color[rgb]{0,0,0}\makebox(0,0)[b]{\smash{\s{B}}}}%
    \put(0.65297076,0.10643316){\color[rgb]{0,0,0}\makebox(0,0)[b]{\smash{\s{A}}}}%
    \put(0.6165417,0.10036139){\color[rgb]{0,0,0}\makebox(0,0)[b]{\smash{\f{f^\dagger}}}}%
    \put(0.68939987,0.10036139){\color[rgb]{0,0,0}\makebox(0,0)[b]{\smash{\f{f}}}}%
    \put(0.53761202,0.10036139){\color[rgb]{0,0,0}\makebox(0,0)[b]{\smash{\f{=}}}}%
    \put(0.76832955,0.10036139){\color[rgb]{0,0,0}\makebox(0,0)[b]{\smash{\f{=}}}}%
    \put(0.16724967,0.10643295){\color[rgb]{0,0,0}\makebox(0,0)[b]{\smash{\s{B}}}}%
    \put(0.24010784,0.10643295){\color[rgb]{0,0,0}\makebox(0,0)[b]{\smash{\s{B}}}}%
    \put(0.42225324,0.10643295){\color[rgb]{0,0,0}\makebox(0,0)[b]{\smash{\s{B}}}}%
    \put(0.49511146,0.10643295){\color[rgb]{0,0,0}\makebox(0,0)[b]{\smash{\s{B}}}}%
    \put(0.31296601,0.10643295){\color[rgb]{0,0,0}\makebox(0,0)[b]{\smash{\s{A}}}}%
    \put(0.2765369,0.03357442){\color[rgb]{0,0,0}\makebox(0,0)[b]{\smash{\f{\sigma^2}}}}%
    \put(0.3858242,0.10036118){\color[rgb]{0,0,0}\makebox(0,0)[b]{\smash{\f{Q}}}}%
    \put(0.2765369,0.10036118){\color[rgb]{0,0,0}\makebox(0,0)[b]{\smash{\f{Q^\dagger}}}}%
    \put(0.45868229,0.10036118){\color[rgb]{0,0,0}\makebox(0,0)[b]{\smash{\f{U}}}}%
    \put(0.20367873,0.10036118){\color[rgb]{0,0,0}\makebox(0,0)[b]{\smash{\f{U^\dagger}}}}%
    \put(0.33118054,0.04571733){\color[rgb]{0,0,0}\makebox(0,0)[b]{\smash{\s{A}}}}%
    \put(0.34939504,0.10643295){\color[rgb]{0,0,0}\makebox(0,0)[b]{\smash{\s{A}}}}%
    \put(0.01546181,0.10643295){\color[rgb]{0,0,0}\makebox(0,0)[b]{\smash{\s{B}}}}%
    \put(0.08224846,0.10643295){\color[rgb]{0,0,0}\makebox(0,0)[b]{\smash{\s{B}}}}%
    \put(0.0458194,0.10036118){\color[rgb]{0,0,0}\makebox(0,0)[b]{\smash{\f{\rho}}}}%
    \put(0.12474906,0.10036139){\color[rgb]{0,0,0}\makebox(0,0)[b]{\smash{\f{=}}}}%
  \end{picture}%
\endgroup%

\ee
The first equality is the spectral decomposition of an arbitrary positive
operator~$\rho$ of rank $\le \dim A$.
The second introduces an arbitrary unitary matrix~$V$ to construct the
SVD~\mbox{$f = U \Sigma V$}.
Finally, the $f$ operators are converted
to the corresponding bipartite states~$\ket{f}$ using cups and caps.
Conversely, any vector~$\ket{f}$ inserted into the diagram on the right yields a positive~$\rho$.
\end{proof}
\end{proposition}

These results can be used to derive correspondence rules
between the properties of operators and the bipartite pure states dual to them, as shown in Table~\ref{tab:opstate}.
\begin{table}[h!]
\caption{Cup-induced isomorphism between operators and bipartite pure quantum states.}
\begin{tabular}{ll}
operator & state vector\\
$f: A \to B$ & $\ket{\opstateup{f}} \in A \otimes B$\\
\hline
$\I$ & $\cupk{}$\\
$\Tr(f)$ & $\braket{\subset}{\opstateup{f}}$\\
real & real\\
symmetric %($f^T = f$)
 & symmetric
($\ket{\opstateup{f}} = \SWAP{}\ket{\opstateup{f}} = \ket{\opstatedown{f}}$)\\
antisymmetric %($f^T = -f$)
 & antisymmetric\\
hermitian %($f^\dagger = f$)
 & $\SWAP{}\ket{\opstateup{f}} = \ket{\cc{\opstateup{f}}}$\\
unitary & locally equivalent to a Bell state\\
% SL(n) & SLOCC-equivalent...
rank & Schmidt rank\\
invertible (full-rank) & full Schmidt rank\\
rank-1 ($f = \ketbra{x}{y}$) & factorizable\\
diagonal & ddot$^{0 \to 2}$\\
%positive & ?\\
%trace-1 & ?\\
%state operator & ?\\
SVD & Schmidt decomposition\\
spectral decomposition & conjugate state decomposition\\
\hline
\end{tabular}
\label{tab:opstate}
\end{table}

\section{Channel-state duality}
\label{sec:channelop}

Much like in the previous section, we now use cups and caps to
construct an isomorphism between linear operator maps (channels)
and bipartite operators, and use it to prove several correspondence rules.
%\todo{say what are they good for}
% numerical implementation of random channels (inherit the distribution from the states)
% analysis of channel properties
% extremals
%

\subsection{Definitions}

Given two Hilbert spaces, $A$ and~$B$, a linear operator
\be
\Omega: \End(A) \to \End(B)
\ee
mapping linear operators on~$A$ to linear operators on~$B$ is called a
\emph{channel} between $A$ and~$B$.
Compatible channels can be concatenated, and all channels of the form
$\Omega: \End(A) \to \End(A)$ form a monoid, i.e. the concatenation
is associative and there is an identity element.
Inverses are not guaranteed, hence not a group.

\begin{definition}[Choi-Jamiołkowski isomorphism]
Using cup- and cap-induced dualities it follows that 
\be
\Hom(\End(A), \End(B)) \isom \Hom(A \otimes A, B \otimes B)
\isom \End(A \otimes B).
\ee
Hence any channel
$\Omega: \End(A) \to \End(B)$
may be represented equally well using
the related linear maps
\begin{align}
\supop{\Omega}&: A \otimes A \to B \otimes B \quad \text{and}\\
\choi{\Omega}&: A \otimes B \to A \otimes B,
\end{align}
presented in \Figref{fig:op_map}a.
$\supop{\Omega}$ is the usual ``vec-superoperator'' representation of~$\Omega$, \ie{}
an operator operating on vectorized operators:
$\ket{\opstateup{\Omega(\rho)}} = \supop{\Omega} \ket{\opstateup{\rho}}$.
It is commonly used in numerical implementations.
The second one, $\choi{\Omega}$, is often called the Choi matrix
of~$\Omega$ and corresponds to $\supop{\Omega}$ with the 
inputs and outputs permuted as shown in~\Figref{fig:op_map}b.
%The permutation corresponds to a a partial transpose followed by a \SWAP{} and another partial transpose.
In these representations, the identity channel is given by
$\supop{\I} = \I$ and $\choi{\I} = \cuprho{}$, respectively.

\begin{figure}[h]
\renewcommand{\fscale}{1.0} \def\svgwidth{0.75\textwidth} 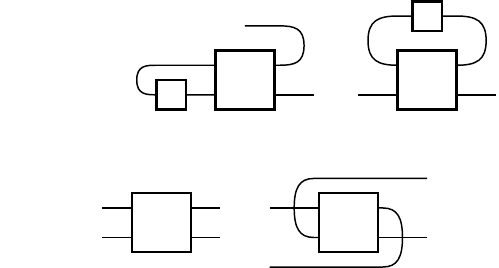
\caption{Channel-operator identities.
(a) Cup-induced isomorphisms between
\mbox{$\Omega \in \Hom(\End(A), \End(B))$},
\mbox{$\supop{\Omega} \in \Hom(A \otimes A, B \otimes B)$}
and \mbox{$\choi{\Omega} \in \End(A \otimes B)$}.
(b) Choi matrix~$\choi{\Omega}$ in terms of the superoperator~$\supop{\Omega}$.
\label{fig:op_map}}
\end{figure}
\end{definition}

We shall now classify different types of channels based on the properties they conserve.
\begin{definition}[Basic properties of channels]
A channel $\Omega$ is
\begin{center}
\begin{tabular}{llll}
(i) & hermitianness-preserving (HP)
& \hspace{1em} iff \hspace{1em}
& $\rho$ is hermitian $\implies \: \Omega(\rho)$ is hermitian\\
(ii) & positivity-preserving (PP)
& \hspace{1em} iff \hspace{1em}
& $\rho \ge 0 \: \implies \: \Omega(\rho) \ge 0$\\
(iii) & completely positivity-preserving (CPP)\footnote{Also known as the complete positivity of operator maps.}
& \hspace{1em} iff \hspace{1em}
& $\I_C \otimes \Omega$ is PP for all~$C$\\
(iv) & trace-preserving (TP)
& \hspace{1em} iff \hspace{1em}
& $\Tr(\Omega(\rho)) = \Tr(\rho)$ for all~$\rho$\\
(v) & unital
& \hspace{1em} iff \hspace{1em}
& $\Omega(\I_A) = \I_B$
\end{tabular}
\end{center}
\end{definition}
If $\Omega$ is HP, this implies that
$\I_C \otimes \Omega $ is HP for all~$C$.
This can be shown by
expanding a bipartite hermitian~$\rho$ in a
factorizable hermitian operator basis, and then
using Lemma~\ref{lemma:omega_tensor_identity_applied}.
However, $\Omega$ is PP does \emph{not} imply that
$\I_C \otimes \Omega $ is PP.
This is why we need to introduce the stronger property,~CPP.
Clearly CPP $\implies$ PP $\implies$ HP. 
%PP implies HP since every hermitian operator can be expressed as the
%difference of two positive operators.
Note that the order of the factors in a tensor product
carries no fundamental importance, hence the identity $\I_C$ could be on
the right as well.

Since $\Tr(\choi{\Omega}) = \Tr(\Omega(\I_A))$, we find that
for a TP channel $\Tr(\choi{\Omega}) = \dim A$, and
for a unital channel $\Tr(\choi{\Omega}) = \dim B$.
Therefore if $\Omega$ is both TP and unital (\emph{doubly stochastic}),
this immediately implies that $\dim A = \dim B$.

\begin{definition}[Quantum channel]
In quantum mechanics, any state can be described using a \emph{state
operator}~$\rho$, also called a density operator, that is positive
and has unit trace. A \emph{quantum channel} is any linear map that maps
state operators to state operators, also when applied only to a part of
a larger system. Therefore it has to be both CPP and TP.
\end{definition}

We will now show an important property of the Choi matrix:
Both concatenation and tensor product of channels correspond to
a tensor product of the corresponding Choi matrices, conjugated by
some additional wire structure.
\begin{definition}[Concatenated channels]
\label{def:concatenated_ch}
Given channels \mbox{$\Omega_1: \End(A) \to \End(B)$} and\\
\mbox{$\Omega_2: \End(B) \to \End(C)$} we can concatenate them:
\mbox{$\Omega = \Omega_2 \circ \Omega_1$}.
% $\supop{\Omega}$ is obtained simply by multiplying the corresponding supops
The Choi matrix $\choi{\Omega}$ is obtained by tensoring the
Choi matrices of the concatenated channels together and then conjugating
with $Q = \I_{A} \otimes \capbra{B} \otimes \I_{C}$:
%to obtain
%$\choi{\Omega} = Q (\choi{\Omega_1} \otimes \choi{\Omega_2}) Q^\dagger$:
\be
\notag
\renewcommand{\fscale}{1.0} \def\svgwidth{0.35\textwidth} %% Creator: Inkscape inkscape 0.48.4, www.inkscape.org
%% PDF/EPS/PS + LaTeX output extension by Johan Engelen, 2010
%% Accompanies image file 'concatenated_channel.pdf' (pdf, eps, ps)
%%
%% To include the image in your LaTeX document, write
%%   \input{<filename>.pdf_tex}
%%  instead of
%%   \includegraphics{<filename>.pdf}
%% To scale the image, write
%%   \def\svgwidth{<desired width>}
%%   \input{<filename>.pdf_tex}
%%  instead of
%%   \includegraphics[width=<desired width>]{<filename>.pdf}
%%
%% Images with a different path to the parent latex file can
%% be accessed with the `import' package (which may need to be
%% installed) using
%%   \usepackage{import}
%% in the preamble, and then including the image with
%%   \import{<path to file>}{<filename>.pdf_tex}
%% Alternatively, one can specify
%%   \graphicspath{{<path to file>/}}
%% 
%% For more information, please see info/svg-inkscape on CTAN:
%%   http://tug.ctan.org/tex-archive/info/svg-inkscape
%%
\begingroup%
  \makeatletter%
  \providecommand\color[2][]{%
    \errmessage{(Inkscape) Color is used for the text in Inkscape, but the package 'color.sty' is not loaded}%
    \renewcommand\color[2][]{}%
  }%
  \providecommand\transparent[1]{%
    \errmessage{(Inkscape) Transparency is used (non-zero) for the text in Inkscape, but the package 'transparent.sty' is not loaded}%
    \renewcommand\transparent[1]{}%
  }%
  \providecommand\rotatebox[2]{#2}%
  \ifx\svgwidth\undefined%
    \setlength{\unitlength}{44.11552563bp}%
    \ifx\svgscale\undefined%
      \relax%
    \else%
      \setlength{\unitlength}{\unitlength * \real{\svgscale}}%
    \fi%
  \else%
    \setlength{\unitlength}{\svgwidth}%
  \fi%
  \global\let\svgwidth\undefined%
  \global\let\svgscale\undefined%
  \makeatother%
  \begin{picture}(1,0.53087293)%
    \put(0,0){\includegraphics[width=\unitlength]{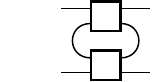}}%
    \put(0.68965128,0.42589151){\color[rgb]{0,0,0}\makebox(0,0)[b]{\smash{\f{\choi{\Omega_1}}}}}%
    \put(0.68965128,0.10461701){\color[rgb]{0,0,0}\makebox(0,0)[b]{\smash{\f{\choi{\Omega_2}}}}}%
    \put(0.2880576,0.2652526){\color[rgb]{0,0,0}\makebox(0,0)[rb]{\smash{\f{\choi{\Omega_2 \circ \Omega_1} =}}}}%
    \put(0.41656776,0.28131704){\color[rgb]{0,0,0}\makebox(0,0)[b]{\smash{\s{B}}}}%
    \put(0.43263151,0.50621041){\color[rgb]{0,0,0}\makebox(0,0)[b]{\smash{\s{A}}}}%
    \put(0.43263151,0.08855257){\color[rgb]{0,0,0}\makebox(0,0)[b]{\smash{\s{C}}}}%
    \put(0.96273562,0.28131704){\color[rgb]{0,0,0}\makebox(0,0)[b]{\smash{\s{B}}}}%
    \put(0.94667201,0.50621041){\color[rgb]{0,0,0}\makebox(0,0)[b]{\smash{\s{A}}}}%
    \put(0.94667201,0.08855257){\color[rgb]{0,0,0}\makebox(0,0)[b]{\smash{\s{C}}}}%
  \end{picture}%
\endgroup%

\ee
\end{definition}

\begin{definition}[Factorizable channel]
\label{def:factorizable_ch}
A channel is factorizable iff it is of the form
\mbox{$\Omega = \Omega_1 \otimes \Omega_2$},
where $\Omega_k: \End(A_k) \to \End(B_k)$.
% $\supop{\Omega}$ is obtained by multiplying the tensor prod with 
% $QQ = \I_{B_1} \otimes \SWAP{}_{B_1 \otimes B_2} \otimes \I_{B_2}$ etc...
In this case $\choi{\Omega}$ is obtained by tensoring the subchannel
Choi matrices and then conjugating
with $Q = \I_{A_1} \otimes \SWAP{}_{B_1 \otimes A_2} \otimes \I_{B_2}$:
%to obtain
%$\choi{\Omega} = Q (\choi{\Omega_1} \otimes \choi{\Omega_2}) Q^\dagger$:
\be
\notag
\renewcommand{\fscale}{1.0} \def\svgwidth{0.4\textwidth} %% Creator: Inkscape inkscape 0.48.4, www.inkscape.org
%% PDF/EPS/PS + LaTeX output extension by Johan Engelen, 2010
%% Accompanies image file 'factorizable_channel.pdf' (pdf, eps, ps)
%%
%% To include the image in your LaTeX document, write
%%   \input{<filename>.pdf_tex}
%%  instead of
%%   \includegraphics{<filename>.pdf}
%% To scale the image, write
%%   \def\svgwidth{<desired width>}
%%   \input{<filename>.pdf_tex}
%%  instead of
%%   \includegraphics[width=<desired width>]{<filename>.pdf}
%%
%% Images with a different path to the parent latex file can
%% be accessed with the `import' package (which may need to be
%% installed) using
%%   \usepackage{import}
%% in the preamble, and then including the image with
%%   \import{<path to file>}{<filename>.pdf_tex}
%% Alternatively, one can specify
%%   \graphicspath{{<path to file>/}}
%% 
%% For more information, please see info/svg-inkscape on CTAN:
%%   http://tug.ctan.org/tex-archive/info/svg-inkscape
%%
\begingroup%
  \makeatletter%
  \providecommand\color[2][]{%
    \errmessage{(Inkscape) Color is used for the text in Inkscape, but the package 'color.sty' is not loaded}%
    \renewcommand\color[2][]{}%
  }%
  \providecommand\transparent[1]{%
    \errmessage{(Inkscape) Transparency is used (non-zero) for the text in Inkscape, but the package 'transparent.sty' is not loaded}%
    \renewcommand\transparent[1]{}%
  }%
  \providecommand\rotatebox[2]{#2}%
  \ifx\svgwidth\undefined%
    \setlength{\unitlength}{55.81335278bp}%
    \ifx\svgscale\undefined%
      \relax%
    \else%
      \setlength{\unitlength}{\unitlength * \real{\svgscale}}%
    \fi%
  \else%
    \setlength{\unitlength}{\svgwidth}%
  \fi%
  \global\let\svgwidth\undefined%
  \global\let\svgscale\undefined%
  \makeatother%
  \begin{picture}(1,0.41960816)%
    \put(0,0){\includegraphics[width=\unitlength]{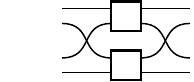}}%
    \put(0.65019675,0.33662962){\color[rgb]{0,0,0}\makebox(0,0)[b]{\smash{\f{\choi{\Omega_1}}}}}%
    \put(0.65019675,0.08269051){\color[rgb]{0,0,0}\makebox(0,0)[b]{\smash{\f{\choi{\Omega_2}}}}}%
    \put(0.24389325,0.20965875){\color[rgb]{0,0,0}\makebox(0,0)[rb]{\smash{\f{\choi{\Omega_1 \otimes \Omega_2} =}}}}%
    \put(0.95492454,0.14617463){\color[rgb]{0,0,0}\makebox(0,0)[b]{\smash{\s{B_1}}}}%
    \put(0.95492454,0.40011462){\color[rgb]{0,0,0}\makebox(0,0)[b]{\smash{\s{A_1}}}}%
    \put(0.95492454,0.06999298){\color[rgb]{0,0,0}\makebox(0,0)[b]{\smash{\s{B_2}}}}%
    \put(0.95492454,0.3239321){\color[rgb]{0,0,0}\makebox(0,0)[b]{\smash{\s{A_2}}}}%
    \put(0.34546907,0.14617288){\color[rgb]{0,0,0}\makebox(0,0)[b]{\smash{\s{B_1}}}}%
    \put(0.34546907,0.40011462){\color[rgb]{0,0,0}\makebox(0,0)[b]{\smash{\s{A_1}}}}%
    \put(0.34546907,0.06999123){\color[rgb]{0,0,0}\makebox(0,0)[b]{\smash{\s{B_2}}}}%
    \put(0.34546907,0.3239321){\color[rgb]{0,0,0}\makebox(0,0)[b]{\smash{\s{A_2}}}}%
  \end{picture}%
\endgroup%

\ee
\end{definition}

\subsection{Correspondence rules}
\label{sec:crules}

Here we present proofs for several well-known theorems connecting
the properties of channels~$\Omega$ to the properties of the
corresponding Choi operators~$\choi{\Omega}$.
As we shall soon see, for every quantum channel the corresponding Choi
operator can be interpreted as the supernormalized state operator of a
bipartite quantum state, and vice versa.
This \emph{channel-state duality} leads to many interesting and useful results.
The diagrammatic approach makes the proofs shorter, more intuitive and easier
to follow.

We shall start by presenting three simple diagrammatic lemmas 
related to tensoring a channel~$\Omega$ with an identity channel.
Remembering that $\choi{\I} = \cuprho{}$ and
inserting it into the diagram in Def.~\ref{def:factorizable_ch}
immediately yields Lemma~\ref{lemma:omega_tensor_identity}.
\begin{lemma}
\label{lemma:omega_tensor_identity}
Choi matrix diagram for the channel $\I_C \otimes \Omega$.
\be
\notag
\renewcommand{\fscale}{1.0} \def\svgwidth{0.75\textwidth} %% Creator: Inkscape inkscape 0.48.4, www.inkscape.org
%% PDF/EPS/PS + LaTeX output extension by Johan Engelen, 2010
%% Accompanies image file 'omega_tensor_identity.pdf' (pdf, eps, ps)
%%
%% To include the image in your LaTeX document, write
%%   \input{<filename>.pdf_tex}
%%  instead of
%%   \includegraphics{<filename>.pdf}
%% To scale the image, write
%%   \def\svgwidth{<desired width>}
%%   \input{<filename>.pdf_tex}
%%  instead of
%%   \includegraphics[width=<desired width>]{<filename>.pdf}
%%
%% Images with a different path to the parent latex file can
%% be accessed with the `import' package (which may need to be
%% installed) using
%%   \usepackage{import}
%% in the preamble, and then including the image with
%%   \import{<path to file>}{<filename>.pdf_tex}
%% Alternatively, one can specify
%%   \graphicspath{{<path to file>/}}
%% 
%% For more information, please see info/svg-inkscape on CTAN:
%%   http://tug.ctan.org/tex-archive/info/svg-inkscape
%%
\begingroup%
  \makeatletter%
  \providecommand\color[2][]{%
    \errmessage{(Inkscape) Color is used for the text in Inkscape, but the package 'color.sty' is not loaded}%
    \renewcommand\color[2][]{}%
  }%
  \providecommand\transparent[1]{%
    \errmessage{(Inkscape) Transparency is used (non-zero) for the text in Inkscape, but the package 'transparent.sty' is not loaded}%
    \renewcommand\transparent[1]{}%
  }%
  \providecommand\rotatebox[2]{#2}%
  \ifx\svgwidth\undefined%
    \setlength{\unitlength}{103.77954218bp}%
    \ifx\svgscale\undefined%
      \relax%
    \else%
      \setlength{\unitlength}{\unitlength * \real{\svgscale}}%
    \fi%
  \else%
    \setlength{\unitlength}{\svgwidth}%
  \fi%
  \global\let\svgwidth\undefined%
  \global\let\svgscale\undefined%
  \makeatother%
  \begin{picture}(1,0.29341043)%
    \put(0,0){\includegraphics[width=\unitlength]{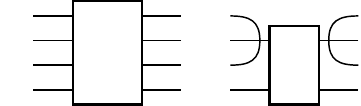}}%
    \put(0.81670585,0.11257431){\color[rgb]{0,0,0}\makebox(0,0)[b]{\smash{\f{\choi{\Omega}}}}}%
    \put(0.57087888,0.13988816){\color[rgb]{0,0,0}\makebox(0,0)[b]{\smash{\f{=}}}}%
    \put(0.29773785,0.15354532){\color[rgb]{0,0,0}\makebox(0,0)[b]{\smash{\f{\choi{\I_C \otimes \Omega}}}}}%
    \put(0.13385318,0.19451634){\color[rgb]{0,0,0}\makebox(0,0)[b]{\smash{\s{A}}}}%
    \put(0.13385318,0.26280167){\color[rgb]{0,0,0}\makebox(0,0)[b]{\smash{\s{C}}}}%
    \put(0.13385318,0.05794566){\color[rgb]{0,0,0}\makebox(0,0)[b]{\smash{\s{B}}}}%
    \put(0.13385318,0.126231){\color[rgb]{0,0,0}\makebox(0,0)[b]{\smash{\s{C}}}}%
    \put(0.46162247,0.19451634){\color[rgb]{0,0,0}\makebox(0,0)[b]{\smash{\s{A}}}}%
    \put(0.46162247,0.26280167){\color[rgb]{0,0,0}\makebox(0,0)[b]{\smash{\s{C}}}}%
    \put(0.46162247,0.05794566){\color[rgb]{0,0,0}\makebox(0,0)[b]{\smash{\s{B}}}}%
    \put(0.46162247,0.126231){\color[rgb]{0,0,0}\makebox(0,0)[b]{\smash{\s{C}}}}%
    \put(0.96693351,0.19451634){\color[rgb]{0,0,0}\makebox(0,0)[b]{\smash{\s{A}}}}%
    \put(0.96693351,0.26280167){\color[rgb]{0,0,0}\makebox(0,0)[b]{\smash{\s{C}}}}%
    \put(0.96693351,0.05794566){\color[rgb]{0,0,0}\makebox(0,0)[b]{\smash{\s{B}}}}%
    \put(0.96693351,0.126231){\color[rgb]{0,0,0}\makebox(0,0)[b]{\smash{\s{C}}}}%
    \put(0.66647831,0.19451634){\color[rgb]{0,0,0}\makebox(0,0)[b]{\smash{\s{A}}}}%
    \put(0.66647831,0.26280167){\color[rgb]{0,0,0}\makebox(0,0)[b]{\smash{\s{C}}}}%
    \put(0.66647831,0.05794566){\color[rgb]{0,0,0}\makebox(0,0)[b]{\smash{\s{B}}}}%
    \put(0.66647831,0.126231){\color[rgb]{0,0,0}\makebox(0,0)[b]{\smash{\s{C}}}}%
  \end{picture}%
\endgroup%

\ee
\end{lemma}

\begin{lemma}
\label{lemma:omega_tensor_identity_applied}
Applying Lemma~\ref{lemma:omega_tensor_identity} to an operator~$\rho$.
\be
\notag
\renewcommand{\fscale}{1.0} \def\svgwidth{0.7\textwidth} %% Creator: Inkscape inkscape 0.48.4, www.inkscape.org
%% PDF/EPS/PS + LaTeX output extension by Johan Engelen, 2010
%% Accompanies image file 'omega_tensor_identity_applied.pdf' (pdf, eps, ps)
%%
%% To include the image in your LaTeX document, write
%%   \input{<filename>.pdf_tex}
%%  instead of
%%   \includegraphics{<filename>.pdf}
%% To scale the image, write
%%   \def\svgwidth{<desired width>}
%%   \input{<filename>.pdf_tex}
%%  instead of
%%   \includegraphics[width=<desired width>]{<filename>.pdf}
%%
%% Images with a different path to the parent latex file can
%% be accessed with the `import' package (which may need to be
%% installed) using
%%   \usepackage{import}
%% in the preamble, and then including the image with
%%   \import{<path to file>}{<filename>.pdf_tex}
%% Alternatively, one can specify
%%   \graphicspath{{<path to file>/}}
%% 
%% For more information, please see info/svg-inkscape on CTAN:
%%   http://tug.ctan.org/tex-archive/info/svg-inkscape
%%
\begingroup%
  \makeatletter%
  \providecommand\color[2][]{%
    \errmessage{(Inkscape) Color is used for the text in Inkscape, but the package 'color.sty' is not loaded}%
    \renewcommand\color[2][]{}%
  }%
  \providecommand\transparent[1]{%
    \errmessage{(Inkscape) Transparency is used (non-zero) for the text in Inkscape, but the package 'transparent.sty' is not loaded}%
    \renewcommand\transparent[1]{}%
  }%
  \providecommand\rotatebox[2]{#2}%
  \ifx\svgwidth\undefined%
    \setlength{\unitlength}{118.30422974bp}%
    \ifx\svgscale\undefined%
      \relax%
    \else%
      \setlength{\unitlength}{\unitlength * \real{\svgscale}}%
    \fi%
  \else%
    \setlength{\unitlength}{\svgwidth}%
  \fi%
  \global\let\svgwidth\undefined%
  \global\let\svgscale\undefined%
  \makeatother%
  \begin{picture}(1,0.31740201)%
    \put(0,0){\includegraphics[width=\unitlength]{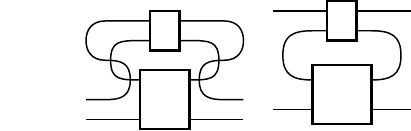}}%
    \put(0.62870479,0.12272508){\color[rgb]{0,0,0}\makebox(0,0)[b]{\smash{\f{=}}}}%
    \put(0.18543293,0.12272508){\color[rgb]{0,0,0}\makebox(0,0)[rb]{\smash{\f{(\I_C \otimes \Omega)(\rho) =}}}}%
    \put(0.83237033,0.08678425){\color[rgb]{0,0,0}\makebox(0,0)[b]{\smash{\f{\choi{\Omega}}}}}%
    \put(0.40107868,0.07480383){\color[rgb]{0,0,0}\makebox(0,0)[b]{\smash{\f{\choi{\Omega}}}}}%
    \put(0.96415388,0.24252801){\color[rgb]{0,0,0}\makebox(0,0)[b]{\smash{\s{A}}}}%
    \put(0.96415388,0.30242969){\color[rgb]{0,0,0}\makebox(0,0)[b]{\smash{\s{C}}}}%
    \put(0.96415388,0.06282341){\color[rgb]{0,0,0}\makebox(0,0)[b]{\smash{\s{B}}}}%
    \put(0.70058678,0.24252801){\color[rgb]{0,0,0}\makebox(0,0)[b]{\smash{\s{A}}}}%
    \put(0.70058678,0.30242969){\color[rgb]{0,0,0}\makebox(0,0)[b]{\smash{\s{C}}}}%
    \put(0.70058678,0.06282341){\color[rgb]{0,0,0}\makebox(0,0)[b]{\smash{\s{B}}}}%
    \put(0.83237033,0.26648926){\color[rgb]{0,0,0}\makebox(0,0)[b]{\smash{\f{\rho}}}}%
    \put(0.40107873,0.24252843){\color[rgb]{0,0,0}\makebox(0,0)[b]{\smash{\f{\rho}}}}%
    \put(0.25731486,0.21856801){\color[rgb]{0,0,0}\makebox(0,0)[b]{\smash{\s{A}}}}%
    \put(0.2453345,0.08678425){\color[rgb]{0,0,0}\makebox(0,0)[b]{\smash{\s{C}}}}%
    \put(0.2453345,0.03886299){\color[rgb]{0,0,0}\makebox(0,0)[b]{\smash{\s{B}}}}%
    \put(0.5448426,0.21856801){\color[rgb]{0,0,0}\makebox(0,0)[b]{\smash{\s{A}}}}%
    \put(0.55682286,0.08678425){\color[rgb]{0,0,0}\makebox(0,0)[b]{\smash{\s{C}}}}%
    \put(0.55682286,0.03886299){\color[rgb]{0,0,0}\makebox(0,0)[b]{\smash{\s{B}}}}%
  \end{picture}%
\endgroup%

\ee
\end{lemma}

\begin{lemma}%[sdfdsf]
\label{lemma:cupstate}
Applying Lemma~\ref{lemma:omega_tensor_identity_applied} in the case where $C = A$
and $\rho = \cuprho{}$,
we obtain $(\I_A \otimes \Omega)(\cuprho{}) = \choi{\Omega}$:
\be
\notag
\renewcommand{\fscale}{1.0} \def\svgwidth{0.7\textwidth} %% Creator: Inkscape inkscape 0.48.4, www.inkscape.org
%% PDF/EPS/PS + LaTeX output extension by Johan Engelen, 2010
%% Accompanies image file 'omega_tensor_identity_cupstate.pdf' (pdf, eps, ps)
%%
%% To include the image in your LaTeX document, write
%%   \input{<filename>.pdf_tex}
%%  instead of
%%   \includegraphics{<filename>.pdf}
%% To scale the image, write
%%   \def\svgwidth{<desired width>}
%%   \input{<filename>.pdf_tex}
%%  instead of
%%   \includegraphics[width=<desired width>]{<filename>.pdf}
%%
%% Images with a different path to the parent latex file can
%% be accessed with the `import' package (which may need to be
%% installed) using
%%   \usepackage{import}
%% in the preamble, and then including the image with
%%   \import{<path to file>}{<filename>.pdf_tex}
%% Alternatively, one can specify
%%   \graphicspath{{<path to file>/}}
%% 
%% For more information, please see info/svg-inkscape on CTAN:
%%   http://tug.ctan.org/tex-archive/info/svg-inkscape
%%
\begingroup%
  \makeatletter%
  \providecommand\color[2][]{%
    \errmessage{(Inkscape) Color is used for the text in Inkscape, but the package 'color.sty' is not loaded}%
    \renewcommand\color[2][]{}%
  }%
  \providecommand\transparent[1]{%
    \errmessage{(Inkscape) Transparency is used (non-zero) for the text in Inkscape, but the package 'transparent.sty' is not loaded}%
    \renewcommand\transparent[1]{}%
  }%
  \providecommand\rotatebox[2]{#2}%
  \ifx\svgwidth\undefined%
    \setlength{\unitlength}{113.1338623bp}%
    \ifx\svgscale\undefined%
      \relax%
    \else%
      \setlength{\unitlength}{\unitlength * \real{\svgscale}}%
    \fi%
  \else%
    \setlength{\unitlength}{\svgwidth}%
  \fi%
  \global\let\svgwidth\undefined%
  \global\let\svgscale\undefined%
  \makeatother%
  \begin{picture}(1,0.33846418)%
    \put(0,0){\includegraphics[width=\unitlength]{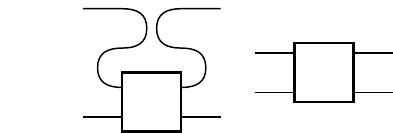}}%
    \put(0.21081221,0.15345742){\color[rgb]{0,0,0}\makebox(0,0)[rb]{\smash{\f{(\I_A\otimes \Omega)(\cuprho) =}}}}%
    \put(0.96248265,0.21609668){\color[rgb]{0,0,0}\makebox(0,0)[b]{\smash{\s{A}}}}%
    \put(0.96248265,0.11587404){\color[rgb]{0,0,0}\makebox(0,0)[b]{\smash{\s{B}}}}%
    \put(0.38620205,0.07829022){\color[rgb]{0,0,0}\makebox(0,0)[b]{\smash{\f{\choi{\Omega}}}}}%
    \put(0.58664743,0.15345742){\color[rgb]{0,0,0}\makebox(0,0)[b]{\smash{\f{=}}}}%
    \put(0.82467631,0.15345742){\color[rgb]{0,0,0}\makebox(0,0)[b]{\smash{\f{\choi{\Omega}}}}}%
    \put(0.68687018,0.21609668){\color[rgb]{0,0,0}\makebox(0,0)[b]{\smash{\s{A}}}}%
    \put(0.68687018,0.11587404){\color[rgb]{0,0,0}\makebox(0,0)[b]{\smash{\s{B}}}}%
    \put(0.24839576,0.32884725){\color[rgb]{0,0,0}\makebox(0,0)[b]{\smash{\s{A}}}}%
    \put(0.24839576,0.05323478){\color[rgb]{0,0,0}\makebox(0,0)[b]{\smash{\s{B}}}}%
    \put(0.52400818,0.32884725){\color[rgb]{0,0,0}\makebox(0,0)[b]{\smash{\s{A}}}}%
    \put(0.52400818,0.05323478){\color[rgb]{0,0,0}\makebox(0,0)[b]{\smash{\s{B}}}}%
  \end{picture}%
\endgroup%

\ee
\end{lemma}
As an immediate consequence of this last lemma,
we obtain an operational interpretation for
the Choi matrix~$\choi{\Omega}$.
It can be understood, up to scaling by $\dim A$, as
the state operator that results when Alice prepares a cup state and
then sends one half of it through the quantum channel~$\Omega$ to Bob.

Next, we will derive three fundamental correspondence rules between
the properties of channels and the corresponding Choi matrices.
\begin{proposition}[Hermitianness preservation~\cite{depillis1967}]
$\Omega$ is HP
iff $\choi{\Omega}$ is hermitian.
\begin{proof}~
\begin{itemize}
\item[$\impliedby$:]
Follows immediately by taking the dagger of the diagram on the
right in \Figref{fig:op_map}a.
\item[$\implies$:]
$\cuprho{}$ is hermitian, which means we may use Lemma~\ref{lemma:cupstate}:\\
\-\hspace{2em} $\Omega$ is HP $\implies$  $\I_A \otimes \Omega$ is HP
$\implies \: (\I_A \otimes \Omega)(\cuprho{}) = \choi{\Omega}$
is hermitian.
\end{itemize}
\end{proof}
\label{th:HP}
\end{proposition}

\begin{proposition}[Positivity preservation]
$\Omega$ is PP iff for all separable
$\sigma \ge 0$ we have $\inprodHS{\sigma}{\choi{\Omega}} \ge 0$.
This also implies that
$\choi{\Omega}$ is hermitian.
\begin{proof}
\begin{align*}
\Omega \: \text{is PP}
\eq \: & \inprodHS{\tau}{\Omega(\varsigma)} \ge 0
\qquad \forall \varsigma, \tau \ge 0
\qquad \text{(by Lemma~\ref{lemma:pos})}\\
\eq \: & \inprodHS{\varsigma^T \otimes \tau}{\choi{\Omega}} \ge 0
\qquad \forall \varsigma, \tau \ge 0 \quad \\
\eq \: & \inprodHS{\sigma}{\choi{\Omega}} \ge 0 \qquad \forall \: \text{separable} \: \sigma \ge 0.
\end{align*}
\end{proof}
\label{th:PP}
\end{proposition}

\begin{proposition}[Complete positivity preservation (Choi's theorem on CPP
maps~\cite{SMR1961,Choi1975})]
\label{th:CPP}
$\Omega$~is CPP
iff $\choi{\Omega}$ is positive.
\begin{proof}~
% tensor prod: a \otimes b \ge 0  \forall a,b \ge 0
% reduction: (\bra{\psi} \otimes \I) a (\ket{\psi} \otimes \I)  \ge 0  \forall a \ge 0
\begin{itemize}
\item[$\impliedby$:]
Choose an arbitrary~$C$,
$\sigma \in \End(C \otimes A) \ge 0$,
$\tau \in \End(C \otimes B) \ge 0$, and
define
\be
\notag
\renewcommand{\fscale}{1.0} \def\svgwidth{0.35\textwidth} %% Creator: Inkscape inkscape 0.48.4, www.inkscape.org
%% PDF/EPS/PS + LaTeX output extension by Johan Engelen, 2010
%% Accompanies image file 'cpp_proof.pdf' (pdf, eps, ps)
%%
%% To include the image in your LaTeX document, write
%%   \input{<filename>.pdf_tex}
%%  instead of
%%   \includegraphics{<filename>.pdf}
%% To scale the image, write
%%   \def\svgwidth{<desired width>}
%%   \input{<filename>.pdf_tex}
%%  instead of
%%   \includegraphics[width=<desired width>]{<filename>.pdf}
%%
%% Images with a different path to the parent latex file can
%% be accessed with the `import' package (which may need to be
%% installed) using
%%   \usepackage{import}
%% in the preamble, and then including the image with
%%   \import{<path to file>}{<filename>.pdf_tex}
%% Alternatively, one can specify
%%   \graphicspath{{<path to file>/}}
%% 
%% For more information, please see info/svg-inkscape on CTAN:
%%   http://tug.ctan.org/tex-archive/info/svg-inkscape
%%
\begingroup%
  \makeatletter%
  \providecommand\color[2][]{%
    \errmessage{(Inkscape) Color is used for the text in Inkscape, but the package 'color.sty' is not loaded}%
    \renewcommand\color[2][]{}%
  }%
  \providecommand\transparent[1]{%
    \errmessage{(Inkscape) Transparency is used (non-zero) for the text in Inkscape, but the package 'transparent.sty' is not loaded}%
    \renewcommand\transparent[1]{}%
  }%
  \providecommand\rotatebox[2]{#2}%
  \ifx\svgwidth\undefined%
    \setlength{\unitlength}{45.80478564bp}%
    \ifx\svgscale\undefined%
      \relax%
    \else%
      \setlength{\unitlength}{\unitlength * \real{\svgscale}}%
    \fi%
  \else%
    \setlength{\unitlength}{\svgwidth}%
  \fi%
  \global\let\svgwidth\undefined%
  \global\let\svgscale\undefined%
  \makeatother%
  \begin{picture}(1,0.51973213)%
    \put(0,0){\includegraphics[width=\unitlength]{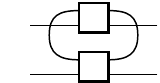}}%
    \put(0.23212949,0.37580296){\color[rgb]{0,0,0}\makebox(0,0)[b]{\smash{\s{A}}}}%
    \put(0.58797045,0.40674589){\color[rgb]{0,0,0}\makebox(0,0)[b]{\smash{\f{\sigma}}}}%
    \put(0.23212949,0.06637693){\color[rgb]{0,0,0}\makebox(0,0)[b]{\smash{\s{B}}}}%
    \put(0.94381114,0.37580296){\color[rgb]{0,0,0}\makebox(0,0)[b]{\smash{\s{A}}}}%
    \put(0.94381114,0.06637693){\color[rgb]{0,0,0}\makebox(0,0)[b]{\smash{\s{B}}}}%
    \put(0.40231424,0.48410266){\color[rgb]{0,0,0}\makebox(0,0)[b]{\smash{\s{C}}}}%
    \put(0.77362639,0.48410266){\color[rgb]{0,0,0}\makebox(0,0)[b]{\smash{\s{C}}}}%
    \put(0.58797045,0.09731985){\color[rgb]{0,0,0}\makebox(0,0)[b]{\smash{\f{\tau}}}}%
    \put(0.0928874,0.25203234){\color[rgb]{0,0,0}\makebox(0,0)[rb]{\smash{\f{\omega =}}}}%
  \end{picture}%
\endgroup%

\ee
Now $\omega \ge 0$, since it is the tensor product of two positive operators
conjugated by some wire structure.
Thus we have
\begin{align*}
\choi{\Omega} \ge 0
\implies \: & \inprodHS{\choi{\Omega}}{\omega} \ge 0 \quad \forall \sigma, \tau \ge 0
\qquad &\text{(by Lemma~\ref{lemma:pos})}\\
\eq \: &\inprodHS{\sigma \otimes \tau}{\choi{\I_C \otimes \Omega}} \ge
0 \quad \forall \sigma, \tau \ge 0
\qquad &\text{(by Lemma~\ref{lemma:omega_tensor_identity})}\\
\eq \: &\I_C \otimes \Omega \: \text{is PP}
\quad \forall C \qquad &\text{(by Proposition~\ref{th:PP})}\\
\eq \: &\Omega \: \text{is CPP}.
\end{align*}
\item[$\implies$:]
$\cuprho{}$ is positive, which means we may use Lemma~\ref{lemma:cupstate}:\\
\-\hspace{2em}
$\Omega$ is CPP $\implies \I_A \otimes \Omega$ is PP
$\implies \: (\I_A \otimes \Omega)(\cuprho{}) = \choi{\Omega}$
is positive.
\end{itemize}
\end{proof}
\end{proposition}

Combining the three propositions above, we obtain
\be
\xymatrix{
  \Omega \text{\: is CPP} \ar@{<=>}[d] \ar@{=>}[r]
& \Omega \text{\: is PP}  \ar@{<=>}[d] \ar@{=>}[r]
& \Omega \text{\: is HP}  \ar@{<=>}[d]\\
\choi{\Omega} \ge 0 \ar@{=>}[r]
& \inprodHS{\sigma}{\choi{\Omega}} \ge 0 \ar@{=>}[r]
& \choi{\Omega}^\dagger = \choi{\Omega}
}
\ee
where $\sigma$ is any separable state operator.

\begin{proposition}[Trace preservation]
\label{th:TP}
$\Omega$ is TP
iff $\Tr_{B}(\choi{\Omega}) = \I_A$.
\begin{proof}
\be
\notag
\renewcommand{\fscale}{1.0} \def\svgwidth{0.9\textwidth} %% Creator: Inkscape inkscape 0.48.4, www.inkscape.org
%% PDF/EPS/PS + LaTeX output extension by Johan Engelen, 2010
%% Accompanies image file 'bistochasticity_TP.pdf' (pdf, eps, ps)
%%
%% To include the image in your LaTeX document, write
%%   \input{<filename>.pdf_tex}
%%  instead of
%%   \includegraphics{<filename>.pdf}
%% To scale the image, write
%%   \def\svgwidth{<desired width>}
%%   \input{<filename>.pdf_tex}
%%  instead of
%%   \includegraphics[width=<desired width>]{<filename>.pdf}
%%
%% Images with a different path to the parent latex file can
%% be accessed with the `import' package (which may need to be
%% installed) using
%%   \usepackage{import}
%% in the preamble, and then including the image with
%%   \import{<path to file>}{<filename>.pdf_tex}
%% Alternatively, one can specify
%%   \graphicspath{{<path to file>/}}
%% 
%% For more information, please see info/svg-inkscape on CTAN:
%%   http://tug.ctan.org/tex-archive/info/svg-inkscape
%%
\begingroup%
  \makeatletter%
  \providecommand\color[2][]{%
    \errmessage{(Inkscape) Color is used for the text in Inkscape, but the package 'color.sty' is not loaded}%
    \renewcommand\color[2][]{}%
  }%
  \providecommand\transparent[1]{%
    \errmessage{(Inkscape) Transparency is used (non-zero) for the text in Inkscape, but the package 'transparent.sty' is not loaded}%
    \renewcommand\transparent[1]{}%
  }%
  \providecommand\rotatebox[2]{#2}%
  \ifx\svgwidth\undefined%
    \setlength{\unitlength}{138.00349084bp}%
    \ifx\svgscale\undefined%
      \relax%
    \else%
      \setlength{\unitlength}{\unitlength * \real{\svgscale}}%
    \fi%
  \else%
    \setlength{\unitlength}{\svgwidth}%
  \fi%
  \global\let\svgwidth\undefined%
  \global\let\svgscale\undefined%
  \makeatother%
  \begin{picture}(1,0.30198511)%
    \put(0,0){\includegraphics[width=\unitlength]{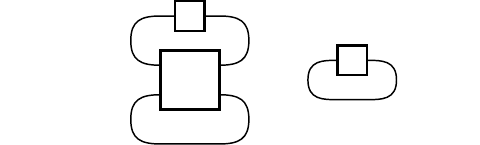}}%
    \put(0.65287852,0.1864055){\color[rgb]{0,0,0}\makebox(0,0)[b]{\smash{\s{A}}}}%
    \put(0.73504011,0.17613522){\color[rgb]{0,0,0}\makebox(0,0)[b]{\smash{\f{\rho}}}}%
    \put(0.57071697,0.13505447){\color[rgb]{0,0,0}\makebox(0,0)[b]{\smash{\f{=}}}}%
    \put(0.27288135,0.14532475){\color[rgb]{0,0,0}\makebox(0,0)[b]{\smash{\s{A}}}}%
    \put(0.39612365,0.13505447){\color[rgb]{0,0,0}\makebox(0,0)[b]{\smash{\f{\choi{\Omega}}}}}%
    \put(0.27288135,0.10424399){\color[rgb]{0,0,0}\makebox(0,0)[b]{\smash{\s{B}}}}%
    \put(0.39612369,0.268567){\color[rgb]{0,0,0}\makebox(0,0)[b]{\smash{\f{\rho}}}}%
    \put(0.23180053,0.13505447){\color[rgb]{0,0,0}\makebox(0,0)[rb]{\smash{\f{\Tr(\Omega(\rho))  =}}}}%
    \put(0.85828245,0.13505447){\color[rgb]{0,0,0}\makebox(0,0)[lb]{\smash{\f{=  \Tr(\rho)}}}}%
  \end{picture}%
\endgroup%

\ee
The above equation presents the TP condition in diagram form.
Denoting $\Tr_{B}(\choi{\Omega})$ by $\omega$, we can see that
\begin{align*}
\Omega \: \text{is TP}
\eq \: & 
\Tr(\Omega(\rho)) = \Tr(\omega^T \rho) = \Tr(\rho) \quad \forall \rho\\
\eq \: & 
\inprodHS{\cc{\omega}}{\rho} = \inprodHS{\I_A}{\rho} \quad \forall \rho\\
\eq \: & 
\inprodHS{\cc{\omega} -\I_A}{\rho} = 0 \quad \forall \rho\\
\eq \: & 
\omega = \Tr_{B}(\choi{\Omega}) = \I_A.
\end{align*}
\end{proof}
By using Lemma~\ref{lemma:span_of_pos_ops}, we can see
that restricting the domain of~$\Omega$ to $\rho \ge 0$ does not
change anything.
\end{proposition}

\begin{proposition}[Unitality]
$\Omega$ is unital
iff $\Tr_{A}(\choi{\Omega}) = \I_B$.
\begin{proof}
Evident by examining the corresponding diagram:
\be
\notag
\renewcommand{\fscale}{1.0} \def\svgwidth{0.9\textwidth} %% Creator: Inkscape inkscape 0.48.4, www.inkscape.org
%% PDF/EPS/PS + LaTeX output extension by Johan Engelen, 2010
%% Accompanies image file 'bistochasticity_unital.pdf' (pdf, eps, ps)
%%
%% To include the image in your LaTeX document, write
%%   \input{<filename>.pdf_tex}
%%  instead of
%%   \includegraphics{<filename>.pdf}
%% To scale the image, write
%%   \def\svgwidth{<desired width>}
%%   \input{<filename>.pdf_tex}
%%  instead of
%%   \includegraphics[width=<desired width>]{<filename>.pdf}
%%
%% Images with a different path to the parent latex file can
%% be accessed with the `import' package (which may need to be
%% installed) using
%%   \usepackage{import}
%% in the preamble, and then including the image with
%%   \import{<path to file>}{<filename>.pdf_tex}
%% Alternatively, one can specify
%%   \graphicspath{{<path to file>/}}
%% 
%% For more information, please see info/svg-inkscape on CTAN:
%%   http://tug.ctan.org/tex-archive/info/svg-inkscape
%%
\begingroup%
  \makeatletter%
  \providecommand\color[2][]{%
    \errmessage{(Inkscape) Color is used for the text in Inkscape, but the package 'color.sty' is not loaded}%
    \renewcommand\color[2][]{}%
  }%
  \providecommand\transparent[1]{%
    \errmessage{(Inkscape) Transparency is used (non-zero) for the text in Inkscape, but the package 'transparent.sty' is not loaded}%
    \renewcommand\transparent[1]{}%
  }%
  \providecommand\rotatebox[2]{#2}%
  \ifx\svgwidth\undefined%
    \setlength{\unitlength}{146.3009104bp}%
    \ifx\svgscale\undefined%
      \relax%
    \else%
      \setlength{\unitlength}{\unitlength * \real{\svgscale}}%
    \fi%
  \else%
    \setlength{\unitlength}{\svgwidth}%
  \fi%
  \global\let\svgwidth\undefined%
  \global\let\svgscale\undefined%
  \makeatother%
  \begin{picture}(1,0.18796876)%
    \put(0,0){\includegraphics[width=\unitlength]{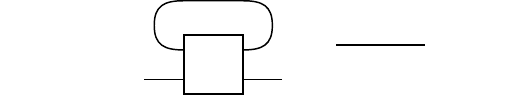}}%
    \put(0.30363579,0.07993879){\color[rgb]{0,0,0}\makebox(0,0)[b]{\smash{\s{A}}}}%
    \put(0.41988847,0.06056319){\color[rgb]{0,0,0}\makebox(0,0)[b]{\smash{\f{\choi{\Omega}}}}}%
    \put(0.5361412,0.04118759){\color[rgb]{0,0,0}\makebox(0,0)[b]{\smash{\s{B}}}}%
    \put(0.30363579,0.04118759){\color[rgb]{0,0,0}\makebox(0,0)[b]{\smash{\s{B}}}}%
    \put(0.60395522,0.09931405){\color[rgb]{0,0,0}\makebox(0,0)[b]{\smash{\f{=}}}}%
    \put(0.26488488,0.09931405){\color[rgb]{0,0,0}\makebox(0,0)[rb]{\smash{\f{\Omega(\I_A) =}}}}%
    \put(0.74927105,0.10900185){\color[rgb]{0,0,0}\makebox(0,0)[b]{\smash{\s{B}}}}%
    \put(0.86552373,0.09931405){\color[rgb]{0,0,0}\makebox(0,0)[lb]{\smash{\f{= \I_B}}}}%
  \end{picture}%
\endgroup%

\ee
\end{proof}
\label{th:unitality}
\end{proposition}

Let us now take a look at a few illustrative examples of channels, quantum and otherwise.

\begin{example}[Information-erasing channel]
The quantum channel given by $\choi{\Omega} = \I_A \otimes \rho_{\text{out}}$,
where $\rho_{\text{out}}$ is a valid quantum state,
is clearly both CPP and TP.
It maps any quantum state to $\rho_{\text{out}}$, thereby erasing all
the information in the input state.
As a special case, for
$\rho_{\text{out}} = \I_B / \dim B$ we obtain the channel
$\choi{\Omega} = \I_{AB} / \dim B$ which maps
any quantum state on~$A$ to the maximum-entropy state on~$B$.
\end{example}

\begin{example}[Unitary channel]
A channel of the form
$\choi{\Omega} = (\I \otimes U) \cuprho{} (\I \otimes U^\dagger)$
is easily seen to be CPP, TP and unital. It corresponds to a quantum
evolution by the unitary propagator~$U$.
% TODO figure
\end{example}

\begin{example}[Transposing channel]
\label{ex:transpose}
Taking the transpose of a state is a classic example of an operation
that is TP, unital and PP but not CPP, and thus not a valid quantum evolution.
This can be shown diagrammatically by forming the Choi
matrix~$\choi{\Omega_{\text{T}}} = \SWAP{}_{A \otimes A}$
(see \Figref{fig:transpose})
and showing that it is not positive by presenting a state~$\ket{\psi}$
that corresponds to a strictly negative eigenvalue.
In this case $\ket{\psi}$ can be chosen to be
any antisymmetric state in $A \otimes A$ (which always exist whenever
$A$ is nontrivial).
However, $\Omega_{\text{T}}$ is easily seen to be both TP and
unital by taking the appropriate partial traces of $\choi{\Omega_{\text{T}}}$,
and PP by
$\inprodHS{\sigma}{\choi{\Omega_{\text{T}}}} = \sum_k p_k \Tr(\varsigma_k \tau_k) \ge 0$,
where the inequality is obtained using Lemma~\ref{lemma:pos_product}.
\begin{figure}[h!]
\renewcommand{\fscale}{1.0} \def\svgwidth{0.45\textwidth} 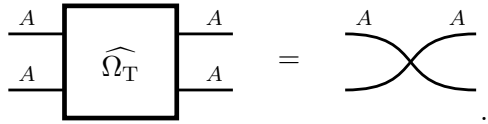.
\caption{Transposing channel.
\label{fig:transpose}
}
\end{figure}
\end{example}

\begin{figure}[h]
\renewcommand{\fscale}{1.0} \def\svgwidth{0.75\textwidth} 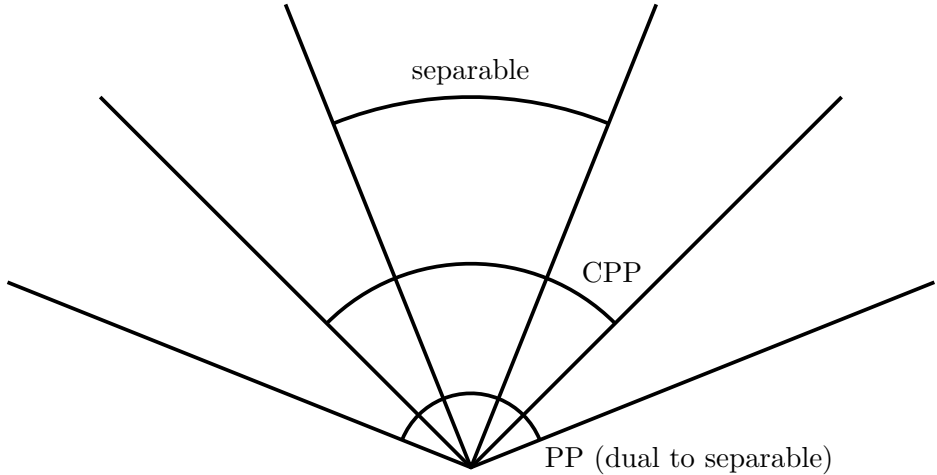
\caption{For any pair of Hilbert spaces $A$ and $B$,
the linear space of Choi matrices
$\choi{\Omega} \in \End(A \otimes B)$
is isomorphic to the linear space of channels
$\Omega: \End(A) \to \End(B)$.
The cone of CPP channels (positive Choi matrices) is self-dual.
The cone of separable channels is dual to the cone of PP channels.
%TODO What about dual \emph{channels}???
\label{fig:choi_space}}
\end{figure}

\Figref{fig:choi_space} illustrates the relationships between some
of the most important subsets of Choi matrices.

\subsection{Concatenated and factorizable channels}

Using the Choi matrix diagrams in Defs.~\ref{def:concatenated_ch}
and~\ref{def:factorizable_ch}
together with the correspondence rules in Sec.~\ref{sec:crules},
we may quickly derive the following properties for concatenated and factorizable channels.
In both cases the Choi matrix has the structure
$\choi{\Omega} = Q (\choi{\Omega_1} \otimes \choi{\Omega_2}) Q^\dagger$,
which means that we may often use similar proofs.
%\mbox{$\Omega = \Omega_1 \otimes \Omega_2$}.
\begin{theorem}[Properties of factorizable channels]~
\begin{itemize}
\item[(i)]
\begin{align*}
\Omega_1, \Omega_2 \quad \text{are HP}
\implies \: &
\choi{\Omega}^\dagger = Q (\choi{\Omega_1}^\dagger \otimes
\choi{\Omega_2}^\dagger) Q^\dagger = \choi{\Omega}\\
\implies \: &
\Omega_2 \circ \Omega_1
\quad \text{and} \quad
\Omega_1 \otimes \Omega_2 \quad \text{are HP}.
\end{align*}
The converse is not true since $i \otimes i = -1$.

\item[(ii)]
\begin{align*}
\Omega_1, \Omega_2 \quad \text{are PP}
\implies \: &
\Omega_1(\rho) \ge 0 \quad \forall \rho \ge 0\\
\implies \: &
(\Omega_2 \circ \Omega_1)(\rho) \ge 0 \quad \forall \rho \ge 0\\
\implies \: &
\Omega_2 \circ \Omega_1 \quad \text{is PP}.
\end{align*}
The converse is not true since $-1 \times -1 = 1$.

Not true for factorizable channels since $\I$ is PP, yet
$\I \otimes \Omega$ is not always PP.
Again, this is the reason we need to introduce the concept of CPP.

\item[(iii)]
% this is done better below
%\begin{align*}
%\Omega \quad \text{is CPP}
%\iff \: &
%\I_C \otimes \Omega \quad \text{is PP} \quad \forall C\\
%\iff \: &
%\I_D \otimes \I_X \otimes \Omega \quad \text{is PP} \quad \forall D,X\\
%\iff \: &
%\I_X \otimes \Omega \quad \text{is CPP} \quad \forall X
%\end{align*}
\begin{align*}
\Omega_1, \Omega_2 \quad \text{are CPP}
\implies \: &
\choi{\Omega} = Q (\choi{\Omega_1} \otimes \choi{\Omega_2}) Q^\dagger \ge 0\\
\implies \: &
\Omega_2 \circ \Omega_1
\quad \text{and} \quad
\Omega_1 \otimes \Omega_2 \quad \text{are CPP}.
\end{align*}
The converse is not true since $-1 \otimes -1 = 1$.

\item[(iv)]
Concatenated channel:
\begin{align*}
\Omega_1, \Omega_2 \quad \text{are TP}
\implies \: &
\Tr_{C}(\choi{\Omega})
= \Tr_{B}(\choi{\Omega_1}) = \I_{A}\\
\implies \: &
\Omega_2 \circ \Omega_1 \quad \text{is TP}.
\end{align*}

Factorizable channel:
\begin{align*}
\Omega_1, \Omega_2 \quad \text{are TP}
\implies \: &
\Tr_{B}(\choi{\Omega})
= \Tr_{B_1}(\choi{\Omega_1}) \otimes \Tr_{B_2}(\choi{\Omega_2}) = \I_{A}\\
\implies \: &
\Omega_1 \otimes \Omega_2 \quad \text{is TP}.
\end{align*}

The converse is not true since $x \I \otimes x^{-1}\I = \I$.

\item[(v)]
Concatenated channel:
\begin{align*}
\Omega_1, \Omega_2 \quad \text{are unital}
\implies \: &
\Tr_{A}(\choi{\Omega})
= \Tr_{B}(\choi{\Omega_2}) = \I_{C}\\
\implies \: &
\Omega_2 \circ \Omega_1 \quad \text{is unital}.
\end{align*}

Factorizable channel:
\begin{align*}
\Omega_1, \Omega_2 \quad \text{are unital}
\implies \: &
\Tr_{A}(\choi{\Omega})
= \Tr_{A_1}(\choi{\Omega_1}) \otimes \Tr_{A_2}(\choi{\Omega_2}) = \I_{B}\\
\implies \: &
\Omega_1 \otimes \Omega_2 \quad \text{is unital}.
\end{align*}
The converse is not true (like above).
\end{itemize}
\end{theorem}

The identity channel $\choi{\I} = \cuprho{}$
is CPP, TP and unital.
This means that we can use these results to obtain the
properties of $\I \otimes \Omega$.

\begin{example}[Partial transpose]
Partial transpose of a bipartite state amounts to
transposing one of the subsystems while leaving the other one
invariant.
It thus corresponds to the channel
$\Omega_{\text{PT}} = \I \otimes \Omega_{\text{T}}$,
where $\Omega_{\text{T}}$ is the transposing channel from Example~\ref{ex:transpose}.
Using the results above, we immediately see that
$\Omega_{\text{PT}}$
is HP, TP and unital, but not PP.
% TODO SHOW! (PPT criterion!).

These properties can also be shown directly by forming the Choi
matrix~$\choi{\Omega}_{\text{PT}}$ (see \Figref{fig:pt}).
\begin{figure}[h!]
\renewcommand{\fscale}{1.0} \def\svgwidth{0.6\textwidth} 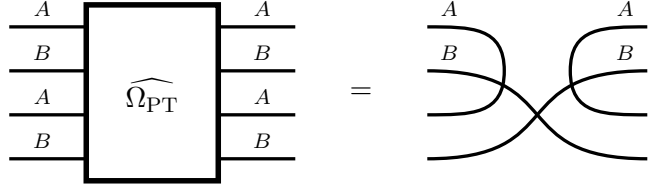
\caption{Partial transpose w.r.t. subsystem~$B$.
\label{fig:pt}
}
\end{figure}
\end{example}

\subsection{Decompositions}

Using the results of Sec.~\ref{sec:crules}, we may now
use standard matrix decompositions to derive corresponding
decompositions for channels.

\begin{figure}[h!]
\renewcommand{\fscale}{1.0} \def\svgwidth{1.0\textwidth} 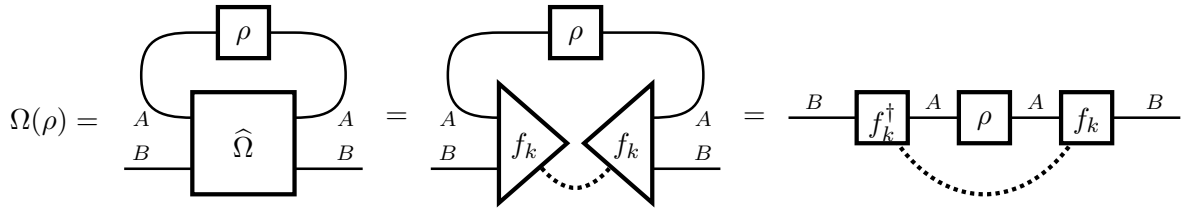
\caption{Spectral decomposition of~$\choi{\Omega}$ is equivalent to the Kraus
  operator sum representation of~$\Omega$. The eigenvalues are nonnegative and have been absorbed into $f_k$
  and~$f_k^\dagger$.
  The dashed line denotes summation over the shared index.
%\yo{Here I would let omega be a diamond, and then take the root of omega and 
%break it into two halves, moving one to the left f and the other to
%the right.  Also, I would remove the dashed lines, since this is
%outside of the formal language that we are using.  I suggest we
%replace the dashed lines by simply putting omega inside a triangle in
%the middle picture, where all sums are replaced with color per the
%Corey convention (see below).}
}
\label{fig:kraus-sd}
\end{figure}

\begin{proposition}[Kraus operator sum representation $\eq$ spectral
decomposition]
Given the Choi matrix $\choi{\Omega} \ge 0$, we may write its spectral decomposition as
\be
\notag
\choi{\Omega} = \sum_k \omega_k \ketbra{\opstateup{\psi_k}}{\opstateup{\psi_k}}
 = \sum_k \ketbra{\opstateup{f_k}}{\opstateup{f_k}},
\ee
where the eigenvalues $\omega_k \ge 0$ can be absorbed into the rescaled eigenvectors
$\ket{\opstateup{f_k}} = \sqrt{\omega_k} \ket{\opstateup{\psi_k}}$.
By elementary diagram manipulation as shown
in~\Figref{fig:kraus-sd}, we can see that this yields the Kraus
operator sum representation for the CPP channel~$\Omega$, and vice versa.
The number of Kraus operators~$f_k$ in the representation is equal to $\Rank \choi{\Omega}$.
Likewise, $\Omega$ is TP if and only if
$\I_A = \Tr_{B}(\choi{\Omega}) = \sum_k f^\dagger_k f_k$,
% $\sum_k \omega_k f^T_k f^*_k$. By transposing this we obtain
the familiar criterion on the completeness of a set of Kraus operators.
% TODO unitary degree of freedom: Ax = Uxy By
% TODO the above spectral decomp is the canonical Kraus decomp
\end{proposition}

%\yo{David Corey came up with the idea of having tensors of the same (non-gray scale) color to be summed over.  I suggest we say that we adopt the Corey convention and 
%use this here since it is somehow more formal than adding anything intuitive into the diagrams to get the point across.}

\begin{comment}
\begin{proposition}[Factorization $\eq$ purification]
$\Omega$ is CPP iff
\begin{proof}
Using Propositions~\ref{th:CPP} and~\ref{th:purification}, we have
$\Omega$ is CPP $\eq$ $\choi{\Omega} \ge 0$ $\eq$
$\choi{\Omega} = \Tr_C(\ketbra{\omega}{\omega})_{A \otimes B \otimes C}$.
%TODO leads to Stinespring theorem...
\end{proof}
\end{proposition}
\end{comment}

\subsection{Dual channels}

\begin{definition}[Dual channel]
Given a channel $\Omega$,
we define its \emph{dual channel}
$\dualc{\Omega}: \End(B) \to \End(A)$
with respect to the Hilbert-Schmidt inner product
$\inprodHS{\cdot}{\cdot}$ to be the channel that fulfills
\be
\inprodHS{\obs}{\Omega(\rho)} =
\inprodHS{\dualc{\Omega}(\obs)}{\rho}
\qquad \forall \: \rho, \obs.
\ee
From the definitions in \Figref{fig:op_map}a
%, together with the reshape invariance of the Hilbert-Schmidt inner
%product (Prop.~\ref{th:hs})
it immediately follows that
\begin{align}
\widetilde{\dualc{\Omega}}
&= \supop{\Omega}^\dagger \qquad \text{and}\\
\label{eq:dualchoi}
\choi{\dualc{\Omega}}
&= \SWAP{}_{A \otimes B}~\cc{\choi{\Omega}}~\SWAP{}_{B \otimes A}.
\end{align}
When $\Omega$ is a quantum channel and $\rho$ a quantum state,
$\obs$ can be interpreted as a hermitian observable or a POVM element, and the
duality transformation itself as the change between the Schrödinger and Heisenberg pictures.
\end{definition}

Using Eq.~\eqref{eq:dualchoi} and the correspondence results in Sec.~\ref{sec:crules},
we may derive a set of equivalences
between the properties of channels and their duals, presented in
Table~\ref{tab:dualchannel}.
\begin{table}[h!]
\caption{Equivalences between the properties of channels and their duals.}
\begin{tabular}{ll}
channel & dual channel\\
$\Omega: \End(A) \to \End(B)$ & $\dualc{\Omega}: \End(B) \to \End(A)$\\
\hline
HP & HP\\
PP & PP\\
CPP & CPP\\
TP & unital\\
unital & TP\\
unitary, $U$ & unitary, $U^\dagger$\\
% Kraus operators: $\{f_k^\dagger\}_k$ & $\{f_k^\dagger\}_k$ ordering?\\
\hline
\end{tabular}
\label{tab:dualchannel}
\end{table}

%TODO self-dual channel?

\begin{table}
\caption{Isomorphism between channels and bipartitite operators.}
\begin{tabular}{ll}
channel & Choi operator\\
$\Omega: \End(A) \to \End(B)$ & $\choi{\Omega}: A \otimes B \to A \otimes B$\\
\hline
HP & hermitian\\
PP & in the dual cone to separable states\\
CPP & positive\\
TP & $\Tr_B(\choi{\Omega}) = \I_A$\\
unital & $\Tr_A(\choi{\Omega}) = \I_B$\\
$\Tr(\Omega(\I_A))$ & $\Tr(\choi{\Omega})$\\
pure, single Kraus term & pure\\
unitary & locally equivalent to a Bell state\\
$\inprodHS{X^T}{\rho} Y$ & factorizable, $X \otimes Y$\\
separable % POVM with specific result states? Holevo form?
& separable\\
%? & maximally entangled\\
%invertible & ?\\
%(maximally) entangling & ?\\
%``scaled-bistochastic''? & scaled isotropic state (\cite{Horodecki1997}?)\\
%unital quantum channel, ? & scaled Werner state $W(\alpha)$\\
%Lindblad-form & ?\\
%projection & ?\\
Kraus decomposition & spectral decomposition\\
Kraus rank & rank\\
%? & Schmidt rank\\
\hline
\end{tabular}
\label{tab:channelop}
\end{table}

%\subsection{Local equivalence}
%Two operations are locally equivalent if one can be converted to the
%other using local operations immediately before and after it.

% using cups/caps show that local equivalence of ops and states is connected

\section{Discussion}
\label{sec:discussion}
We summarize the results we have presented on channel-operator correspondence in Table~\ref{tab:channelop}.

Using tensor network diagrams to describe quantum mechanical systems
is a broad topic with has attracted considerable interest recently. 
Reference~\cite{wood2011} gives detailed diagrammatic
proofs of several results not elaborated here, e.g. the Stinespring theorem.

\begin{acknowledgments}
Parts of this work were completed with financial support from the Fondazione Compagnia di San Paolo
through the Q-ARACNE project.
\end{acknowledgments}

\bibliography{qc}

%merlin.mbs apsrev4-1.bst 2010-07-25 4.21a (PWD, AO, DPC) hacked
%Control: key (0)
%Control: author (0) dotless jnrlst
%Control: editor formatted (1) identically to author
%Control: production of article title (0) allowed
%Control: page (1) range
%Control: year (0) verbatim
%Control: production of eprint (0) enabled
\begin{thebibliography}{11}%
\makeatletter
\providecommand \@ifxundefined [1]{%
 \@ifx{#1\undefined}
}%
\providecommand \@ifnum [1]{%
 \ifnum #1\expandafter \@firstoftwo
 \else \expandafter \@secondoftwo
 \fi
}%
\providecommand \@ifx [1]{%
 \ifx #1\expandafter \@firstoftwo
 \else \expandafter \@secondoftwo
 \fi
}%
\providecommand \natexlab [1]{#1}%
\providecommand \enquote  [1]{``#1''}%
\providecommand \bibnamefont  [1]{#1}%
\providecommand \bibfnamefont [1]{#1}%
\providecommand \citenamefont [1]{#1}%
\providecommand \href@noop [0]{\@secondoftwo}%
\providecommand \href [0]{\begingroup \@sanitize@url \@href}%
\providecommand \@href[1]{\@@startlink{#1}\@@href}%
\providecommand \@@href[1]{\endgroup#1\@@endlink}%
\providecommand \@sanitize@url [0]{\catcode `\\12\catcode `\$12\catcode
  `\&12\catcode `\#12\catcode `\^12\catcode `\_12\catcode `\%12\relax}%
\providecommand \@@startlink[1]{}%
\providecommand \@@endlink[0]{}%
\providecommand \url  [0]{\begingroup\@sanitize@url \@url }%
\providecommand \@url [1]{\endgroup\@href {#1}{\urlprefix }}%
\providecommand \urlprefix  [0]{URL }%
\providecommand \Eprint [0]{\href }%
\providecommand \doibase [0]{http://dx.doi.org/}%
\providecommand \selectlanguage [0]{\@gobble}%
\providecommand \bibinfo  [0]{\@secondoftwo}%
\providecommand \bibfield  [0]{\@secondoftwo}%
\providecommand \translation [1]{[#1]}%
\providecommand \BibitemOpen [0]{}%
\providecommand \bibitemStop [0]{}%
\providecommand \bibitemNoStop [0]{.\EOS\space}%
\providecommand \EOS [0]{\spacefactor3000\relax}%
\providecommand \BibitemShut  [1]{\csname bibitem#1\endcsname}%
\let\auto@bib@innerbib\@empty
%</preamble>
\bibitem [{\citenamefont {Penrose}(1971)}]{Penrose_string_diagram}%
  \BibitemOpen
  \bibfield  {author} {\bibinfo {author} {\bibfnamefont {Roger}\ \bibnamefont
  {Penrose}},\ }\bibfield  {title} {\enquote {\bibinfo {title} {Applications of
  negative dimensional tensors},}\ }in\ \href@noop {} {\emph {\bibinfo
  {booktitle} {Combinatorial Mathematics and its Applications}}},\ \bibinfo
  {editor} {edited by\ \bibinfo {editor} {\bibfnamefont {D.}~\bibnamefont
  {Welsh}}}\ (\bibinfo  {publisher} {Academic Press, New York},\ \bibinfo
  {year} {1971})\ pp.\ \bibinfo {pages} {221--244}\BibitemShut {NoStop}%
\bibitem [{\citenamefont {Arrighi}\ and\ \citenamefont
  {Patricot}(2004)}]{arrighi2004}%
  \BibitemOpen
  \bibfield  {author} {\bibinfo {author} {\bibfnamefont {P.}~\bibnamefont
  {Arrighi}}\ and\ \bibinfo {author} {\bibfnamefont {C.}~\bibnamefont
  {Patricot}},\ }\bibfield  {title} {\enquote {\bibinfo {title} {On quantum
  operations as quantum states},}\ }\href {\doibase 10.1016/j.aop.2003.11.005}
  {\bibfield  {journal} {\bibinfo  {journal} {Annals of Physics}\ }\textbf
  {\bibinfo {volume} {311}},\ \bibinfo {pages} {26--52} (\bibinfo {year}
  {2004})},\ \Eprint {http://arxiv.org/abs/quant-ph/0307024}
  {arXiv:quant-ph/0307024} \BibitemShut {NoStop}%
\bibitem [{\citenamefont {Verstraete}\ and\ \citenamefont
  {Verschelde}(2003)}]{vv2003}%
  \BibitemOpen
  \bibfield  {author} {\bibinfo {author} {\bibfnamefont {Frank}\ \bibnamefont
  {Verstraete}}\ and\ \bibinfo {author} {\bibfnamefont {Henri}\ \bibnamefont
  {Verschelde}},\ }\href@noop {} {\enquote {\bibinfo {title} {On quantum
  channels.}}\ } (\bibinfo {year} {2003}),\ \Eprint
  {http://arxiv.org/abs/quant-ph/0202124} {arXiv:quant-ph/0202124} \BibitemShut
  {NoStop}%
\bibitem [{\citenamefont {Wood}\ \emph {et~al.}(2015)\citenamefont {Wood},
  \citenamefont {Biamonte},\ and\ \citenamefont {Cory}}]{wood2011}%
  \BibitemOpen
  \bibfield  {author} {\bibinfo {author} {\bibfnamefont {Christopher~J.}\
  \bibnamefont {Wood}}, \bibinfo {author} {\bibfnamefont {Jacob~D.}\
  \bibnamefont {Biamonte}}, \ and\ \bibinfo {author} {\bibfnamefont {David~G.}\
  \bibnamefont {Cory}},\ }\bibfield  {title} {\enquote {\bibinfo {title}
  {Tensor networks and graphical calculus for open quantum systems},}\
  }\href@noop {} {\bibfield  {journal} {\bibinfo  {journal} {Quant. Inf. and
  Comp.}\ }\textbf {\bibinfo {volume} {15}},\ \bibinfo {pages} {0759--0811}
  (\bibinfo {year} {2015})},\ \Eprint {http://arxiv.org/abs/1111.6950}
  {arXiv:1111.6950} \BibitemShut {NoStop}%
\bibitem [{\citenamefont {Selinger}(2011)}]{Selinger09}%
  \BibitemOpen
  \bibfield  {author} {\bibinfo {author} {\bibfnamefont {Peter}\ \bibnamefont
  {Selinger}},\ }\bibfield  {title} {{\enquote
  {\bibinfo {title} {A survey of graphical languages for monoidal
  categories},}\ }}in\ \href {\doibase 10.1007/978-3-642-12821-9_4}
  {{\emph {\bibinfo {booktitle} {New Structures for
  Physics}}}},\ \bibinfo {series} {Lecture Notes in Physics}, Vol.\ \bibinfo
  {volume} {813},\ \bibinfo {editor} {edited by\ \bibinfo {editor}
  {\bibfnamefont {Bob}\ \bibnamefont {Coecke}}}\ (\bibinfo  {publisher}
  {Springer Berlin Heidelberg},\ \bibinfo {year} {2011})\ pp.\ \bibinfo {pages}
  {289--355},\ \Eprint {http://arxiv.org/abs/0908.3347} {arXiv:0908.3347}
  \BibitemShut {NoStop}%
\bibitem [{\citenamefont {Bergholm}\ and\ \citenamefont
  {Biamonte}(2011)}]{BB10}%
  \BibitemOpen
  \bibfield  {author} {\bibinfo {author} {\bibfnamefont {Ville}\ \bibnamefont
  {Bergholm}}\ and\ \bibinfo {author} {\bibfnamefont {Jacob~D.}\ \bibnamefont
  {Biamonte}},\ }\bibfield  {title} {\enquote {\bibinfo {title} {Categorical
  quantum circuits},}\ }\href {\doibase 10.1088/1751-8113/44/24/245304}
  {\bibfield  {journal} {\bibinfo  {journal} {J. Phys. A: Math. Theor.}\
  }\textbf {\bibinfo {volume} {44}},\ \bibinfo {pages} {245304} (\bibinfo
  {year} {2011})},\ \Eprint {http://arxiv.org/abs/1010.4840} {arXiv:1010.4840}
  \BibitemShut {NoStop}%
\bibitem [{\citenamefont {Biamonte}\ \emph {et~al.}(2013)\citenamefont
  {Biamonte}, \citenamefont {Bergholm},\ and\ \citenamefont
  {Lanzagorta}}]{BBL2012}%
  \BibitemOpen
  \bibfield  {author} {\bibinfo {author} {\bibfnamefont {Jacob~D.}\
  \bibnamefont {Biamonte}}, \bibinfo {author} {\bibfnamefont {Ville}\
  \bibnamefont {Bergholm}}, \ and\ \bibinfo {author} {\bibfnamefont {Marco}\
  \bibnamefont {Lanzagorta}},\ }\bibfield  {title} {\enquote {\bibinfo {title}
  {Tensor network methods for invariant theory},}\ }\href {\doibase
  10.1088/1751-8113/46/47/475301} {\bibfield  {journal} {\bibinfo  {journal}
  {J. Phys. A: Math. Theor.}\ }\textbf {\bibinfo {volume} {46}},\ \bibinfo
  {pages} {475301} (\bibinfo {year} {2013})},\ \Eprint
  {http://arxiv.org/abs/1209.0631} {arXiv:1209.0631} \BibitemShut {NoStop}%
\bibitem [{Note1()}]{Note1}%
  \BibitemOpen
  \bibinfo {note} {Also known as the complete positivity of operator
  maps.}\BibitemShut {Stop}%
\bibitem [{\citenamefont {de~Pillis}(1967)}]{depillis1967}%
  \BibitemOpen
  \bibfield  {author} {\bibinfo {author} {\bibfnamefont {John~Emanuel}\
  \bibnamefont {de~Pillis}},\ }\bibfield  {title} {\enquote {\bibinfo {title}
  {Linear transformations which preserve hermitian and positive semidefinite
  operators},}\ }\href {\doibase 10.2140/pjm.1967.23.129} {\bibfield  {journal}
  {\bibinfo  {journal} {Pacific J. Math.}\ }\textbf {\bibinfo {volume} {23}},\
  \bibinfo {pages} {129--137} (\bibinfo {year} {1967})}\BibitemShut {NoStop}%
\bibitem [{\citenamefont {Sudarshan}\ \emph {et~al.}(1961)\citenamefont
  {Sudarshan}, \citenamefont {Mathews},\ and\ \citenamefont {Rau}}]{SMR1961}%
  \BibitemOpen
  \bibfield  {author} {\bibinfo {author} {\bibfnamefont {E.~C.~G.}\
  \bibnamefont {Sudarshan}}, \bibinfo {author} {\bibfnamefont {P.~M.}\
  \bibnamefont {Mathews}}, \ and\ \bibinfo {author} {\bibfnamefont
  {Jayaseetha}\ \bibnamefont {Rau}},\ }\bibfield  {title} {\enquote {\bibinfo
  {title} {Stochastic dynamics of quantum-mechanical systems},}\ }\href
  {\doibase 10.1103/PhysRev.121.920} {\bibfield  {journal} {\bibinfo  {journal}
  {Phys. Rev.}\ }\textbf {\bibinfo {volume} {121}},\ \bibinfo {pages}
  {920--924} (\bibinfo {year} {1961})}\BibitemShut {NoStop}%
\bibitem [{\citenamefont {Choi}(1975)}]{Choi1975}%
  \BibitemOpen
  \bibfield  {author} {\bibinfo {author} {\bibfnamefont {Man-Duen}\
  \bibnamefont {Choi}},\ }\bibfield  {title} {\enquote {\bibinfo {title}
  {Completely positive linear maps on complex matrices},}\ }\href {\doibase
  10.1016/0024-3795(75)90075-0} {\bibfield  {journal} {\bibinfo  {journal}
  {Linear Algebra and its Applications}\ }\textbf {\bibinfo {volume} {10}},\
  \bibinfo {pages} {285--290} (\bibinfo {year} {1975})}\BibitemShut {NoStop}%
\end{thebibliography}%

\clearpage 
\appendix 
\section{Some linear algebra background}
\label{sec:linalg}

Throughout this work we explicitly use the Hilbert-Schmidt inner product and the Frobenius norm.
\begin{definition}[Hilbert-Schmidt inner product]
\label{def:hs}
Let $f, g: A \to B$ be linear operators between the Hilbert spaces $A$ and~$B$.
The Hilbert-Schmidt inner product on the space~$\Hom(A,B)$ is defined as
\be
\inprodHS{f}{g} := \Tr(f^\dagger g).
\ee
It induces the Frobenius norm:
\be
\norm{f}_F := \sqrt{\inprodHS{f}{f}}.
\ee
For kets and bras interpreted as operators in $\Hom(\eye, A)$
and $\Hom(A, \eye)$, respectively,
the Hilbert-Schmidt inner product reduces to the usual
inner product of vectors on a Hilbert space.
\end{definition}

\begin{definition}[Dual cone]
Given a linear space~$A$ and the Hilbert-Schmidt inner product on it, we define the \emph{dual cone} of a set $Q \subset A$
to be
\be
Q^* = \{y \in A | \inprodHS{y}{x} \ge 0 \quad \forall x \in Q \}.
\ee
$Q^*$ is clearly a convex cone, since any linear combination of its elements with nonnegative scalar multipliers yields another element of~$Q^*$.
%It is also easy to see that $Q^{**}$ is the convex hull of~$Q$ scaled into a cone???
% $Q \subset Q^{**}$.
% and $Q^* = Q^{***}$.
We say that $Q$ is \emph{self-dual} iff $Q = Q^*$.
\end{definition}

\begin{proposition}[Leg-bending is a HS isometry]
\label{th:hs}
Let $f, g: A \to B$ be linear maps between the Hilbert spaces $A$ and~$B$.
Since bending of tensor legs using cups and caps amounts
to merely reshaping the corresponding matrix, any leg-bending operation~$\Xi$
preserves the Hilbert-Schmidt inner product:
\be
\inprodHS{\Xi(f)}{\Xi(g)}
= \Tr\left((\Xi(f))^\dagger \Xi(g)\right)
= \Tr(f^\dagger g)
= \inprodHS{f}{g}.
\ee
This is easy to verify using diagrams.
Hence any such~$\Xi$ is an isometry and preserves orthogonality
between sets of tensors.
The induced Frobenius norm is also preserved.
\end{proposition}

\begin{lemma}[]
\label{lemma:span_of_pos_ops}
On a complex Hilbert space~$A$,
% NOTE on a real H space (d = 2 for example), antisymmetric matrices cannot be written as scalar*symmetric.
% We get every symmetric matrix as a difference of positive matrices, but there's no way to get antisymmetric ones out of them.
% Easy to numerically verify by Gram-Schmidting random positive matrices.
positive operators span all linear operators:
\be
\Span \, \{\rho \in \End(A)| \rho \ge 0\} = \End(A).
\ee
\begin{proof}~
Any operator $x \in \End(A)$ can be decomposed to its hermitian and antihermitian parts: $x = h_1 +ih_2$,
where $h_1, h_2$ are hermitian.
Furthermore, any hermitian operator can be expressed as the difference of two positive operators:
$h = \rho_1-\rho_2$.
\end{proof}
Corollary:
Assume we have an arbitrary linear functional $\phi: \End(A) \to \C$.
Now
\be
\phi(\rho) = 0 \quad \forall \rho \ge 0
\qquad \iff \qquad
\phi(x) = 0 \quad \forall x
\qquad \iff \qquad
\phi = 0,
\ee
and thus
\be
\inprodHS{x}{\rho} = 0 \quad \forall \rho \ge 0
\qquad \iff \qquad
x = 0,
\ee
since $\inprodHS{x}{\cdot}$ is a linear functional.
\end{lemma}
%NOTE Because of this lemma one cannot obtain anything new by restricting
%the treatment to just positive operators instead of all ops later on.

\begin{lemma}[]
\label{lemma:pos_product}
% weaker result: eigendecomposition: $\Tr(\sigma \tau) = \sum_k \sigma_k \bra{\psi_k} \tau \ket{\psi_k} \ge 0$.
The product of two positive operators $\sigma$, $\tau$ is not
necessarily positive
itself, but it has a nonnegative spectrum of eigenvalues.
\begin{proof}
Given $\sigma$, $\tau \ge 0$, assume $\lambda$ is an eigenvalue of $\sigma \tau$:
\be
\notag
\sigma \tau \ket{x} = \lambda \ket{x} \quad \text{for some} \: \ket{x} \neq 0.
\ee
This implies
\be
\notag
\sqrt{\tau} \sigma \sqrt{\tau} (\sqrt{\tau} \ket{x}) = \lambda (\sqrt{\tau} \ket{x}).
\ee
If $\sqrt{\tau} \ket{x} \neq 0$ this means that $\lambda$ is an eigenvalue of
$\sqrt{\tau} \sigma \sqrt{\tau} \ge 0$, and thus nonnegative.
If $\sqrt{\tau} \ket{x} = 0$, we have $\ket{x} \in \Ker(\tau)$ and thus
$\lambda = 0$.
\end{proof}
Corollary:
For any two $\sigma$, $\tau \ge 0$ we have
$\inprodHS{\sigma}{\tau} = \Tr(\sigma \tau) \ge 0$.
\end{lemma}

Lemma~\ref{lemma:pos} shows that the set of positive operators is its own dual cone.

% positive ops form a 90-degree cone
\begin{lemma}[]
\label{lemma:pos}
Positive operators form a self-dual convex cone:
\be
\inprodHS{\sigma}{\tau} \ge 0 \quad \forall \sigma \ge 0
\qquad \eq \qquad \tau \ge 0.
\ee
\begin{proof}~
\begin{itemize}
\item[$\impliedby$:]
By Lemma~\ref{lemma:pos_product}.
\item[$\implies$:]
Choosing $\sigma = \ketbra{\psi}{\psi}$ we obtain
$\inprodHS{\sigma}{\tau} = \Tr(\ketbra{\psi}{\psi} \tau)
= \bra{\psi}\tau\ket{\psi}\ge 0$.
Since this holds for an arbitrary~$\ket{\psi}$, we find that $\tau \ge 0$.
% NOTE that the above can only hold if $\tau$ is hermitian, so that's checked too
\end{itemize}
\end{proof}
\end{lemma}

\section{Quantum states, separability and entanglement}
\label{sec:states}

In quantum mechanics, every physical system is associated with a
complex Hilbert space~\HH, called a state space.
Every state of the system can be described using a state
operator~$\rho \in \End(\HH)$, also called a density operator, which is
positive semidefinite and has unit trace, and vice versa.
Given~\HH, the set of all state operators of this system is denoted as~S(\HH),
and is seen to be convex and closed.
It is also bounded due to the unit trace requirement,
without which we would obtain the convex cone of positive operators instead.

The extremal points of $S(\HH)$ constitute the \emph{pure states}~$P(\HH)$, i.e., states of the
form $\rho = \ketbra{\psi}{\psi}$ where $\ket{\psi} \in \HH$.
However, all the boundary points need not be pure states.
This can be seen using a dimensional argument: using $d := \dim_\C \HH$,
clearly $\dim_\R P(\HH) = 2d -2$ whereas
$\dim_\R \partial S(\HH) = d^2 -2$.
In the single-qubit case ($d=2$) $\partial S(\HH)$ coincides with $P(\HH)$, but
in higher dimensions $\dim_\R P(\HH)$ is strictly smaller than $\dim_\R \partial S(\HH)$.

A system is \emph{bipartite} if it has a state space of the
form~$\HH = A \otimes B$, where $A$ and $B$ are the state spaces of
its subsystems (neither of which has to be elementary in the physical sense).
This idea of dividing a system into two parts enables
us to introduce the concept of entanglement.

\begin{definition}[Factorizable, separable and entangled states]
\label{def:separable}
A bipartite state operator $\rho~\in~S(A \otimes B)$
is
\begin{itemize}
\item
\emph{factorizable} iff $\rho = \varsigma \otimes \tau$,
where $\varsigma \in S(A)$ and $\tau \in S(B)$,

\item
\emph{separable} iff it can be expressed as
a convex combination of factorizable state operators:
\be
\rho = \sum_k p_k \: \varsigma_k \otimes \tau_k,
\quad \text{where} \: p_k \ge 0 \: \text{and} \: \sum_k p_k = 1,
\ee
\item
\emph{entangled} iff it is not separable.
\end{itemize}
\end{definition}
Due to its construction, the set of separable states is also convex,
closed and bounded. Again, if one relaxes the unit trace requirement, one
obtains the convex cone of separable positive operators.

\end{document}